\documentclass[]{elsarticle}

\usepackage{lineno,hyperref}
\modulolinenumbers[5]
\usepackage{epsfig} 
\usepackage{graphicx}
\usepackage{caption}
\usepackage{subcaption}
\usepackage{multirow}
\usepackage{multicol}
\usepackage{etoolbox}
\usepackage{setspace}
\usepackage{lipsum}
\usepackage{amssymb,mathtools}
\usepackage{lscape}

\usepackage[utf8]{inputenc}

\makeatletter
\def\ps@pprintTitle{%
 \let\@oddhead\@empty
 \let\@evenhead\@empty
 \def\@oddfoot{}%
 \let\@evenfoot\@oddfoot}
\makeatother
\usepackage{siunitx}
\usepackage[nocfg]{nomencl}
\usepackage{nomencl}
\usepackage{ifthen}
\renewcommand\nomgroup[1]{%
  \ifthenelse{\equal{#1}{A}}{%
    \item[\textbf{Acronyms}]}{
  \ifthenelse{\equal{#1}{R}}{%
    \item[\textbf{Roman Symbols}]}{
  \ifthenelse{\equal{#1}{G}}{%
    \item[\textbf{Greek Symbols}]}{
  \ifthenelse{\equal{#1}{S}}{%
    \item[\textbf{Superscripts}]}{
  \ifthenelse{\equal{#1}{U}}{%
    \item[\textbf{Subscripts}]}{
  \ifthenelse{\equal{#1}{X}}{%
    \item[\textbf{Other Symbols}]}{
  {}}}}}}}}

\makenomenclature
\makenomenclature

\usepackage{fancyhdr}
\makeatletter
\let\runauthor\@author
\let\runtitle\@title
\makeatother

\begin{document}

\begin{frontmatter}

\title{Investigation of  Thermal Adiabatic Boundary Condition on Semitransparent Wall in Combined Radiation and Natural Convection}
\author[]{G Chanakya}
\author[]{Pradeep Kumar\thanks{corresponding author}}



\ead{pradeepkumar@iitmandi.ac.in}

\address{Numerical Experiment Laboratory \\ 
	 (Radiation and Fluid Flow Physics)\\
	School of Engineering\\
	Indian Institute of Technology Mandi\\
	Mandi, Himachal Pradesh, India 175075\\}
\date{}

\begin{abstract}
Two thermal adiabatic boundary conditions arise on the semitransparent window owing to the fact that whether semitransparent window allows the energy to leave the system by radiation mode of heat transfer. It is assumed that being low conductivity of semitransparent material, energy does not leave by conduction mode of heat transfer. This does mean that the semitransparent window may behave as only conductively adiabatic $(q_c=0)$ or combinedly conductively and radiatively adiabatic $(q_c+q_r=0)$. In the present work, the above two thermal adiabatic boundary conditions have been investigated in natural convection problem for the Rayleigh number (Ra) $10^5$ and Prandtl number(Pr) 0.71 in a cavity, whose left vertical wall has been divided into upper and lower parts in the ratio of 4:6. The upper section is semitransparent window, while lower section is isothermal wall at a temperature of 296K. A collimated beam is irradiated with different values (0, 100, 500 and 1000 $W/m^2$) on the semitransparent window at an angle of $45^0$. The cavity is heated from the bottom by convective heating with free stream temperature of 305K and heat transfer coefficient of 50 $W/m^2 K$ while right wall is also isothermal at same temperature as of lower left wall and upper wall is adiabatic. All walls are opaque for radiation except semitransparent window. The results reveal that the dynamics of both the vortices inside the cavity change drastically with irradiation values and also with the boundary conditions on the semitransparent window. The temperature and Nusselt number increase multifold inside the cavity for combinedly conductively and radiatively adiabatic condition than the only conductively adiabatic condition on the semitransparent window.

\end{abstract}
\newpage
\begin{keyword}
\texttt{Semitransparent window; Natural convection; Collimated beam; Symmetrical cooling; Irradiation; Bottom Heating;}
\end{keyword}

\end{frontmatter}

\linenumbers
            
    \nomenclature[a]{$a$}{Co-efficient}
    \nomenclature[a]{$C_{p}$}{Specific heat capacity ($J/kg-K$)}
    \nomenclature[a]{$g$}{Acceleration due to gravity ($m/s^2$)}
    \nomenclature[a]{$G$}{Irradiation ($W/m^2$)}
    \nomenclature[a]{$H$}{Height ($m$)}
    \nomenclature[a]{$I$}{Intensity ($W/m^2$)}
    \nomenclature[a]{$I_{b}$}{Black body intensity ($W/m^2$)}
    \nomenclature[a]{$k$}{Thermal conductivity ($W/mK$)}
    \nomenclature[a]{$L$}{Length of the domain of study ($m$)}
    \nomenclature[a]{$Nu$}{Nusselt number}
    \nomenclature[a]{$p$}{Pressure ($N/m^2$)}
    \nomenclature[a]{$Pr$}{Prandtl number}
	\nomenclature[a]{$q$}{Flux ($W/m^2$)}
	\nomenclature[a]{$Ra$}{Rayleigh number}
	\nomenclature[a]{$u,v$}{Velocity ($m$)}
    \nomenclature[g]{$\beta_{T}$}{Thermal expansion coefficient ($1/K$)}
    \nomenclature[g]{$\epsilon$}{Emissivity}
	\nomenclature[g]{$\kappa_{a}$}{Absorption coefficient ($1/m$)} 
	\nomenclature[g]{$\rho$}{Density of the fluid ($kg/m^3$)}
    \nomenclature[g]{$\tau$}{Optical thickness}
	\nomenclature[g]{$\phi$}{Scalar}
    \nomenclature[U]{$C$}{Conduction}	
	\nomenclature[U]{$c$}{Cold wall}
	\nomenclature[U]{$co$}{Collimated beam}
	\nomenclature[U]{$conv$}{Convection}
	\nomenclature[U]{$f$}{Face centre}
	\nomenclature[U]{$free$}{Free stream}
	\nomenclature[U]{$i,j$}{Tensor indices}
	\nomenclature[U]{$nb$}{Neighbour cell}
	\nomenclature[U]{$p$}{Cell centre}		
	\nomenclature[U]{$R$}{Radiation}
	\nomenclature[U]{$ref$}{Reference}
	\nomenclature[U]{$t$}{Total}
	\nomenclature[U]{$w$}{Wall}

\printnomenclature

\section{Introduction}

The knowledge of heat transfer in buoyancy driven flows is crucial in many natural and engineering applications, like HVAC systems, electronic cooling. In recent years, there has been increasing interest of the researchers in the analysis of multi-physics problems involving all modes of heat transfer.

Many researchers \cite{Torrance,Calcagani,Ganzorolli,Aydin,Sathiyamoorthy} investigated the buoyancy driven flows in 2D enclosures heating from bottom and cooled either one side or both sides. Acharya and Goldstein \cite{Acharya} studied the effect of internal energy sources on the natural convection in an inclined square box. Osman et al. \cite{Osman} investigated the effect of non-newtonian fluid (i.e Bingham fluid) on natural convection, it was observed that Nusselt number was smaller in Bingham fluids than Newtonian fluid for the same Rayleigh and Prandtl numbers. A comprehensive review on natural convection has been performed by Rahimi et al. \cite{Rahimi} and Das et al. \cite{Das}.

The interaction of radiation with natural convection has been investigated by Mondal and Mishra \cite{Bittagopal}, Liu et al. \cite{Liu} and Kumar and Eswaran \cite{Kumar} in a 2-dimensional cavity with different optical thicknesses and reported that there was an considerable effect of radiation on the natural convection. The square cavity has been selected in first two works whereas slanted cavity was used by Kumar and Eswaran. Lari et al. \cite{Lari} also showed that the radiation has considerable effects on the natural convection even at low operating temperatures. Further studies on the effect of different geometrical shapes, like square, triangle and circle at the centre of the square cavity on the combined radiation and natural convection were performed by Mezrhab et al. \cite{Mezrhab},  Sun et al. \cite{Hua} and Mukul et al. \cite{Mukul}, whereas, the study with circle inside the circular cavity has been performed by Xu et al. \cite{Xu}. 

The discreate ordinate (DO) method \cite{Chai} was most acceptable and popular method to solver radiative heat transfer equation before finite volume discreate ordinate method (fvDOM) \cite{Raithby,Chui} and now the researchers mostly use fvDOM in the radiation field. The performance analysis of various methods for radiation transfer equation (RTE) like FVM, DOM, P1, SP3 and P3 have studied by Sun et al. \cite{Yujia}, and reported that the P1 method consumed minimal time than the other methods, however, it suffer in accuracy at low optical thickness of medium whereas FVM gave better accuracy compared to DOM but took more computation time. Assessment of natural convection with radiation for different aspect ratios ranging from 1 to 6, and range of Rayleigh numbers $1.60 \times 10^5$ to $4.67 \times 10^7$ in rectangular enclosure have been studied numerically in \cite{Hakan} and heat transfer correlations has been formulated. It has been reported that the mean Nusselt number increased to 73.35 from 23.63 for aspect ratio 1 to 6. 

Webb and Viskanta \cite{Webb} performed an experiment to study the natural convection with water as working fluid in an rectangular enclosure irradiated with collimated beam from a side and kept  opposite wall on constant temperature, whereas other walls were kept adiabatic. The results showed that there was formation of hydrodynamic boundary layers at the vertical walls. The authors have also analysed numerically the same problem for the prediction of internal heating in the fluid.

Anand and Mishra \cite{Anand} investigated the radiative transfer equation for variable refractive index for participating media. Ben and Dez \cite{Ben} provided an exact expression for the radiative flux for the emitting and absorbing semitransparent medium which has a linear refractive index variation. The above researchers mostly considered the combined diffuse radiation with natural convection, however, little work is available on collimated radiation. Apparently to the present authors knowledge there is no much work focusing the interaction of collimated beam radiation with natural convection. 

Also, the walls of the geometries in all above works have been considered as opaque which does not allow any energy enter into domain through radiation mode of heat transfer. Whereas, semitransparent wall allows energy to enter into the system by radiative mode of heat transfer but may or may not allow to leave the system. This arises two heat flux boundary conditions on the semitransparent window and such conditions are mostly encounter in practical scenarios. In the present work two heat flux boundary conditions on the semitransparent wall for natural convection problem including radiation have been investigated. This paper is outlined as follows; section 2 describes the problem statement followed by mathematical modelling and numerical schemes in section 3. The validation and grid independent test are explained in section 4 and 5, respectively. Section 6 elaborates the results and discussion.Finally, the conclusions of the present numerical study  are provided in section 7.  

\section{Problem statement}

\begin{figure}[t]
    \centering
    \includegraphics[width=6cm]{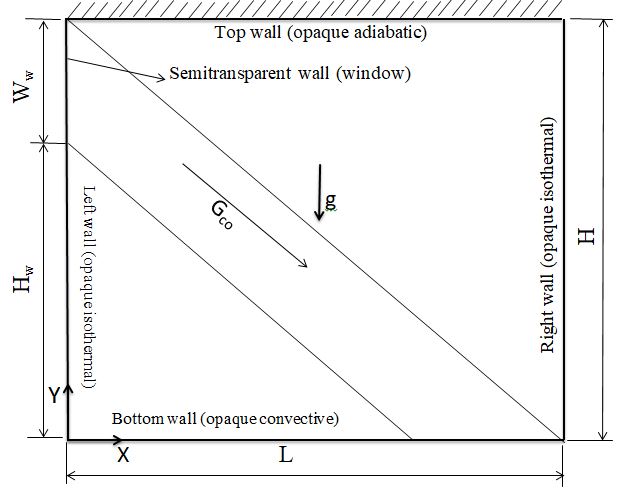}
    \caption{Schematic diagram of cavity for the present problem with collimated beam incident at an angle of $45^0$}
    \label{semi_prob}
\end{figure}

Figure \ref{semi_prob} depicts the geometry for present study. The ecludian co-ordinate axis are along horizontal and vertical walls of cavity and origin is at lower left corner of the cavity. The gravity force acts in negative Y direction. The left wall of the cavity has been divided into two upper and lower parts in the ratio of 4:6, where the lower part of the cavity is opaque while upper part is semitransparent (for the radiation energy) and the rest walls are considered to be opaque. The thermal conditions on the lower left and right vertical walls are constant temperature at 296 K while upper wall is thermally adiabatic. The cavity is heated from bottom by convective heat transfer with heat transfer coefficient of 50 $W/m^2 K$ and free stream temperature is 305 K. The combined radiation and natural convection are considered inside the cavity. The working fluid of the cavity is considered to be transparent for radiation energy and flow is governed buoyancy force corresponds to Rayleigh number $10^5$ and Prandtl number (Pr=0.71). A collimated beam enters through upper left wall inside the cavity at an angle of $45^0$. This semitransparent wall may or may not allow the radiation energy to leave the cavity assuming being low conductivity of semitransparent material, the energy does not leave the system through conduction mode of heat transfer. Based on these two scenarios two thermal conditions on the semitransparent wall are considered

case (A). Conductively Adiabatic: The semitransparent wall allows to leave energy through radiation mode but not by conduction mode, i.e $q_c=0$,

case (B) Combinedly conductively and radiatively adiabatic wall: The semitransparent wall does not allow the energy leave neither by conduction nor by radiation heat transfer, i.e $q_c+q_r=0$.

The effect of above two boundary conditions on the fluid flow and heat transfer characteristics inside the cavity has been studied for irradiation values of 0, 100, 500 and 1000 $W/m^2$.

\section{Mathematical formulation and Numerical procedures}
\subsection{Mathematical formulation}
The following assumptions have been considered for the mathematical modelling of the problem; 
\begin{enumerate}
    \item Flow is steady, laminar, incompressible and two dimensional.
    \item Flow is driven by buoyancy force that is modeled by Boussinesq approximation.
    \item The thermophysical properties of fluid are constant.
    \item The fluid may or may not participate in radiative heat transfer.
    \item The refractive index of the medium is constant and equal to one.
    \item The fluid absorbs and emits but does not scatter the radiation energy.
    \item The transmissivity of semitransparent window is one and zero for other walls.
\end{enumerate}
Based on the above assumptions the governing equations in the Cartesian coordinate system are given by  

\begin{equation} \label{mass:equN} 
\frac{\partial u_i}{\partial x_i} = 0  
\end{equation}
\vspace{-1cm}
\begin{equation} \label{momentum}
\frac{\partial  u_i u_j}{\partial x_j}=-\frac{1}{\rho}\frac{\partial p}{\partial x_i} + \nu\frac{\partial^2u_i}{\partial x_j\partial x_j}+g \beta_{T}(T-T_{c})\delta_{i2}  
\end{equation}

\begin{equation}\label{energy}
\frac{\partial u_jT}{\partial x_j} = \frac{k}{\rho C_p}\frac{\partial^2T}{\partial x_j\partial x_j} - \frac{1}{\rho C_p} \frac{\partial q_{R}}{\partial x_i} 
\end{equation}
\vspace{2cm}
\noindent where $\frac{\partial q_{R_{i}}}{\partial x_{i}}$ is the divergence of the radiative flux, which can be calculated as
\vspace{-1cm}
\begin{eqnarray}
\frac{\partial q_{R_{i}}}{\partial x_{i}}=\kappa_{a}(4\pi I_b-G)\label{div_eq}
\end{eqnarray}
where $\kappa_{a}$ is the absorption coefficient, $I_b$ is the black body intensity and $G$ is the irradiation which is evaluated by integrating the radiative intensity ($I$) in all directions, i.e.,
\begin{eqnarray}
G=\int_{4\pi} I d\Omega
\end{eqnarray}
The intensity field inside the cavity can be obtained by solving the following radiative transfer equation (RTE) 
 
 \begin{equation} 
\label{equN:radiation_1}
\frac{\partial I(\bf \hat{r},\bf \hat{s})}{\partial s}=\kappa_{a}I_{b}(\bf\hat{s})-(\kappa_{a}) \text{I} (\bf\hat{r},\bf \hat{s})
\end{equation}
Where $\bf \hat{r},\bf \hat{s}$ is position and direction vectors, whereas s is path length. The Navier-Stokes equation and temperature equations are subjected to boundary conditions 
\begin{enumerate}
 \item[]\textit{Flow boundary condition}
    \item[] Cavity walls: \hspace{0.2cm} $u_i$=0
 \item[] \textit{Thermal boundary conditions}
    \item[1.] Left wall  (a) Lower part: Isothermal
    \newline \text{\hspace{4.5cm}} T=296 K and 
  \newline  \text{\hspace{0.75cm}}  (b) Upper part: either $q_c=0$ or
  \newline \text{\hspace{4.5cm}} $q_c+q_r=0$
    \item[2.] Right wall at \textit{x}=1;\hspace{0.2cm} T=296K
    \item[3.] Bottom wall \textit{y}=0 ;\hspace{0.2cm} $q_{conv}= h_{free} (T_{free}-T_{w})$
    \item[4.] Top wall at \textit{y}=1; \hspace{0.2cm} $q_{c}+q_{r}=0$
    \item[]where $q_c=-k \frac{\partial T}{\partial n}$ and $ q_{r}=\int_{4\pi}I({\bf r_w})({\bf \hat{n}\cdot\hat{s}})\mathrm{d}\Omega $
\end{enumerate}

\begin{figure}[t]
 \begin{subfigure}{6cm}
    \centering\includegraphics[width=5cm]{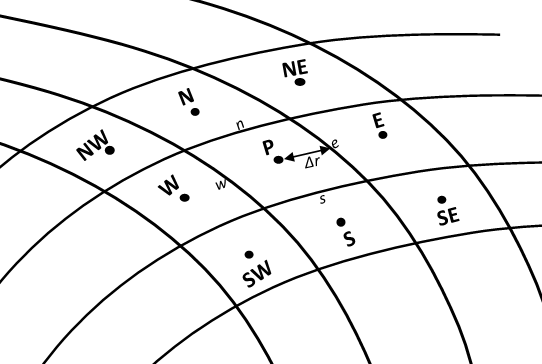}
    \caption{}
    \label{2d_fvm}
  \end{subfigure}\hfill
   \begin{subfigure}{8cm}
    \centering\includegraphics[width=5cm]{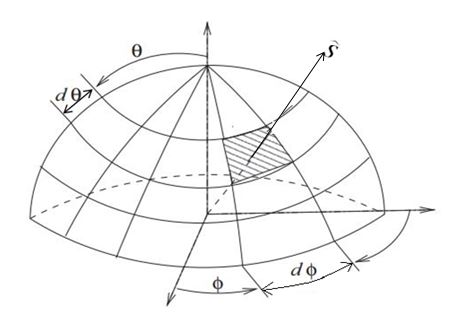}
    \caption{ }
    \label{angular}
  \end{subfigure}\hfill
  \caption{Pictorial representation of (a) cell arrangement for finite volume method for partial differential equation  and (b) Angular discretization for the radiative transfer equation}
\label{diffuse}
\end{figure}

The radiative transfer equation (\ref{equN:radiation_1}) is subjected to following boundary condition, all cavity wall (assumed black wall) except semitransparent window 

\begin{eqnarray}
\noindent I({\bf r_w,\hat{s}})=\epsilon_w I_b({\bf r_w})+\frac{1-\epsilon_w}{\pi}\int_{\bf \hat{n}\cdot\hat{s}>0}I({\bf r_w,\hat{s}})|{\bf \hat{n}\cdot\hat{s}}|\mathrm{d}\Omega.\nonumber\\
\mbox{for}~~{\bf \hat{n}\cdot\hat{s}}<0 \label{rte:bound2}
\end{eqnarray}

where $\hat{n}$ is the surface normal and the emissivity of all walls is considered to be 1.

The semitransparent window is subjected to collimated irradiation $(G_{co})$ of value 1000 $W/m^2$. The boundary condition for RTE on semitransparent window is 
\begin{figure}[t]
 \begin{subfigure}{7cm}
    \centering\includegraphics[width=4cm]{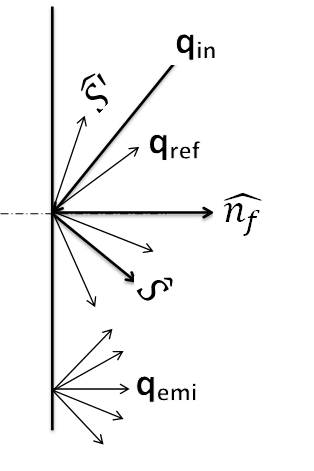}
    \caption{}
    \label{Diff_BC}
  \end{subfigure}\hfill
   \begin{subfigure}{7cm}
    \centering\includegraphics[width=4cm]{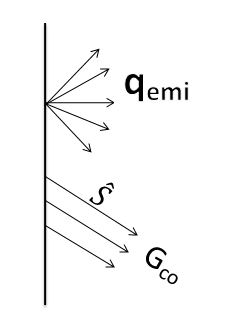}
    \caption{}
    \label{semi_colli}
  \end{subfigure}\hfill
  \caption{Pictorial representation of (a) Diffuse reflection of an incident ray and diffuse emission due to wall temperature (b) Diffuse emission and collimated transmission from a semitransparent wall}
\label{Colli_diffuseBC}
\end{figure}
\begin{eqnarray}
\noindent I({\bf r_w,\hat{s}})=I_{co}({\bf r_w,\hat{s}}) \delta (\theta-45^{0})+\epsilon_w I_b({\bf r_w})+\frac{1-\epsilon_w}{\pi}\int_{\bf \hat{n}\cdot\hat{s}>0}I({\bf r_w,\hat{s}})|{\bf \hat{n}\cdot\hat{s}}|\mathrm{d}\Omega.\nonumber\\
\mbox{for}~~{\bf \hat{n}\cdot\hat{s}}<0 \label{rte:bound3}
\end{eqnarray}
where $\delta (\theta-45^{0})$ is Dirac-delta function,
\begin{equation}
  \delta(\theta-45^{0})=
    \begin{cases}
            1, &         \text{if} \text{\hspace{0.2cm}} \theta=45^{0},\\
            0, &         \text{if} \text{\hspace{0.2cm}} \theta \neq 45^{0}.
    \end{cases}  
\end{equation}

$I_{co}$ is intensity of collimated irradiation and calculated from the irradiation value as below
\begin{equation}
    I_{co}=\frac{G_{co}}{\mathrm{d}\Omega}
\end{equation}
Where $d\Omega$ is the collimated beam width. In the current work, the solid angle of discretized angular space (Fig. \ref{angular}) in collimated direction is considered as beam width of the collimated beam. The pictorial representation of  diffuse emission and reflection and collimated beam radiation from the wall is shown in Fig. \ref{Colli_diffuseBC}. The collimated feature has been developed in OpenFOAM framework an open source software and coupled with other fluid and heat transfer libraries. The combined application have been used for numerical simulation. The OpenFOAM uses the finite volume method (FVM) to solve the Navier-Stokes and the energy equations. The FVM integrates an equation over a control volume (Fig. \ref{2d_fvm}) to convert the partial differential equation into a set of algebraic equations in the form
\begin{equation}
    a_{p}\phi_{p}= \sum_{nb}a_{nb}\phi_{nb}+S
\end{equation}
where $\phi_{p}$ is any scalar and $a_{p}$ is central coefficient, $a_{nb}$ coefficients of neighbouring cells and S is the source values, whereas, RTE eq (\ref{equN:radiation_1}) is converted into a set of algebraic equations by double integration over a control volume and over a control angle. The set of algebraic equations are solved by Preconditioned bi-conjugate gradient (PBiCG) and the details of the algorithm can be found in the book by Patankar \cite{patankar} and Moukalled \cite{Moukalled}. In the present simulation, linear upwind scheme which is second order accurate has been used to interpolate face centred value. The linear upwind scheme is given mathematically as
\begin{equation}
   \phi_{f}=
    \begin{cases}
            \phi_{p}+ \nabla\phi \cdot \nabla r, &     \text{if}  \text{\hspace{0.2cm}}  f_{\phi} > 0,\\
            \phi_{nb}+ \nabla\phi \cdot \nabla r, &     \text{if} \text{\hspace{0.2cm}}  f_{\phi} < 0.
    \end{cases}   
\end{equation}
and $f_{\phi}$ is the flux of the scalar $\phi$ on a face (Fig \ref{2d_fvm}).

 \subsection{Non-dimensional Parameters}
The OpenFOAM simulation produces the results in dimensional quantities. To explain the results in more general form, the simulated results are converted into non-dimensional parameters. The scales for length, velocity, temperature, and conductive and radiative fluxes are L, u$_o$, (T$_{free}$-T$_{c}$), $\kappa$(T$_{free}$-T$_{c}$)/L and $\sigma$T$_{free}^{4}$ respectively, where $u_{o}=\sqrt {L g \beta (T_{free}-T_{c})}$ is convective velocity scale.

The non-dimensional quantities and parameters involved in the present problem are as follows,
\begin{eqnarray}
U =\frac{u}{u_{o}},  \hspace{0.5cm}   V=\frac{v}{u_{o}}, \hspace{0.5cm}  \hspace{0.5cm} X =\frac{x}{L},  \hspace{0.5cm}   Y=\frac{y}{L}, \hspace{0.5cm} \theta =\frac{T-T_{c}}{T_{free}-T_{c}}   
\end{eqnarray}

\begin{eqnarray}
Ra=\frac{g \beta (T_{free}-T_{c})L^{3}}{\nu \alpha}, \hspace{0.5cm}  Pr = \frac{\nu}{\alpha} 
\end{eqnarray}

The optical thickness is defined as $\tau=\kappa_{a} L$ and non-dimensional irradiation is given as
\begin{eqnarray}
\overline{G}=\frac{G}{\sigma T^{4}_{free}} \hspace{1cm}
\end{eqnarray}

The Nu$_{C}$ and Nu$_{R}$ are conductive and radiative Nusselt numbers respectively and defined as
\begin{eqnarray}
Nu_{C}=\frac{q_{Cw}L}{k(T_{free}-T_{c})}, \hspace{1cm} Nu_{R}=\frac{q_{Rw}L}{k(T_{free}-T_{c})} \hspace{1cm} 
\end{eqnarray}
Thus, the total Nusselt number is defined as below,
\begin{eqnarray}
 Nu=Nu_{C}+Nu_{R} \hspace{0.5cm}  \hspace{0.5cm}  
\end{eqnarray}

\section{Validation}

\begin{figure}[!t]
 \begin{subfigure}{6cm}
    \centering\includegraphics[width=7cm]{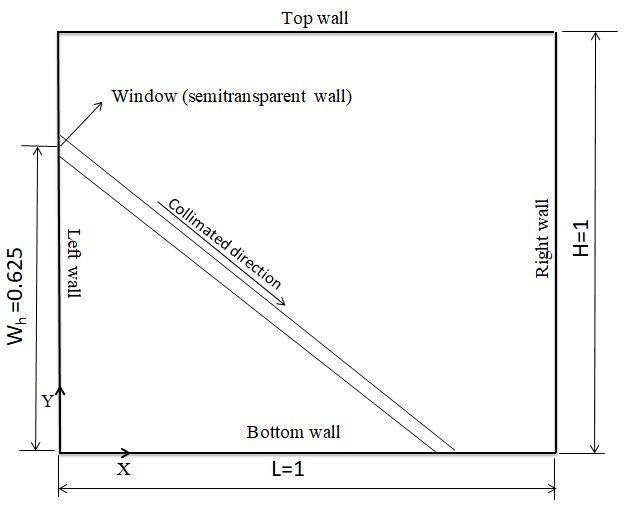}
    \caption{}
    \label{test_collimated}
  \end{subfigure}\hfill
   \begin{subfigure}{8.3cm}
    \centering\includegraphics[width=6cm]{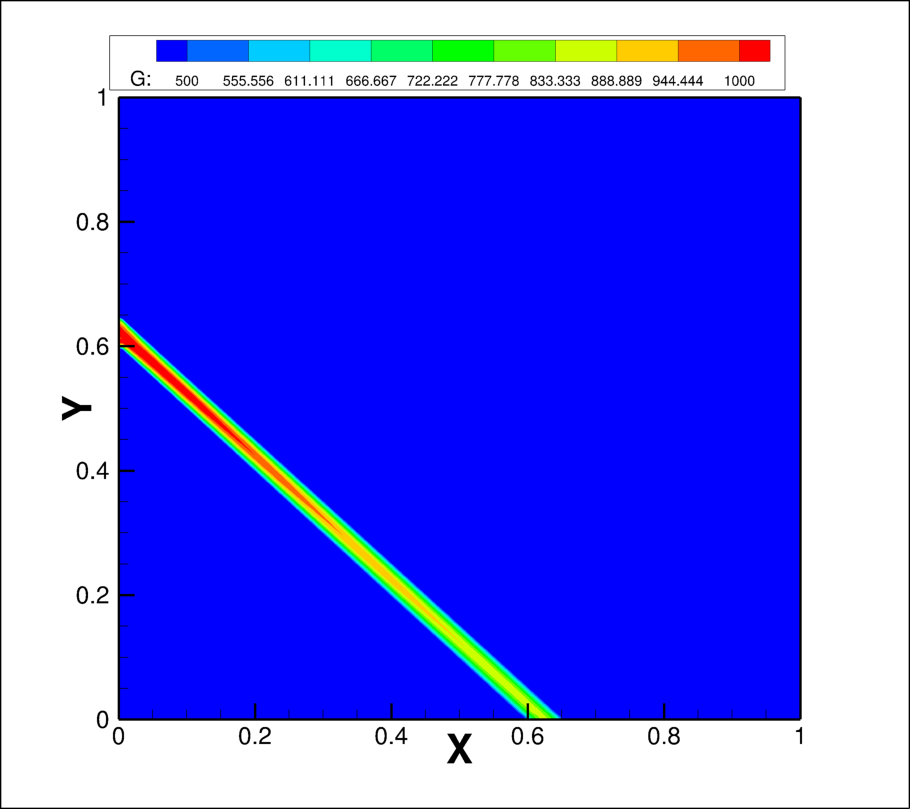}
    \caption{}
    \label{valid_collimated}
  \end{subfigure}\hfill
  \caption{ Schematic for the (a) Test case for the collimated beam radiation and (b) Contour representing the collimated beam travelling normal to the wall}
\label{diffuse}
\end{figure}

\begin{figure}[!htb]
	\centering
	\includegraphics[width=6cm,scale=1]{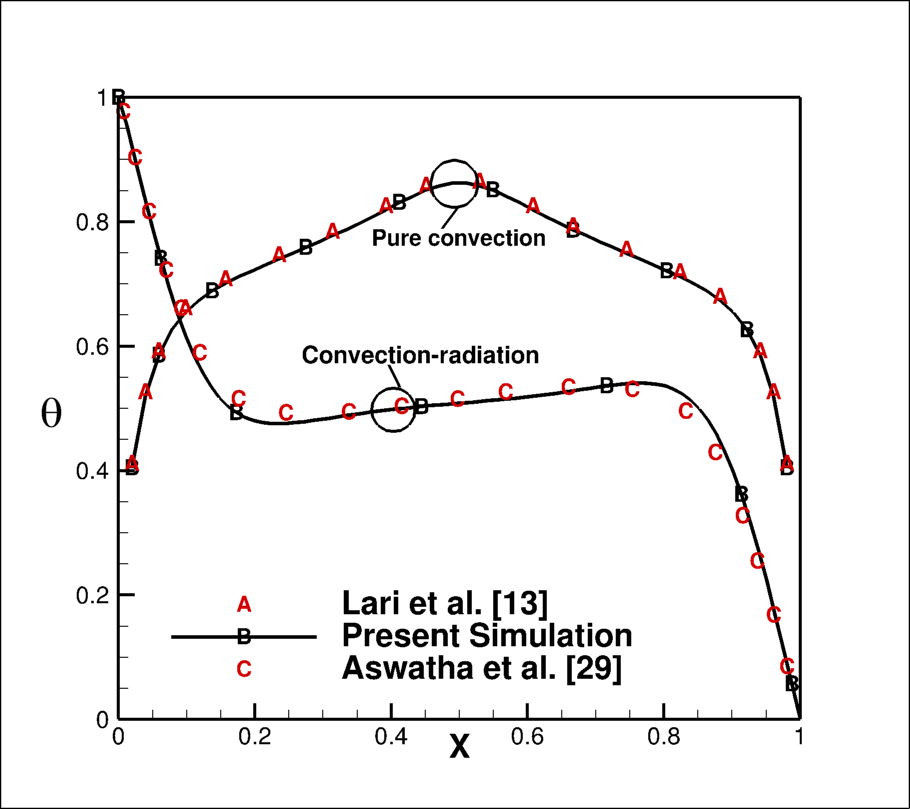}
	\caption{Validation results for pure convection and combined diffuse radiation with natural convection}
	\label{valid_temp}
	
\end{figure}

In the absence of any standard benchmark test case for the present problem, the validation has been performed in three steps, first, the standalone feature of collimated beam irradiation problem, in second step, pure natural convection problem which is heated from the bottom and in the third step combined convection and radiation in defferentially heated cavity have been verified. The collimated irradiation feature \cite {RAD19} has been tested in a square cavity as shown in the Fig. (\ref{test_collimated}). The left side of the wall has s small window of non-dimensional size 0.05 at a non-dimensional height of 0.6. The walls of square cavity are black and cold and also medium is non-participating. A collimated beam is irradiated on the window in $45^0$ direction. It is expected that the beam would travel in oblique direction of $45^0$ angle without any attenuation and hit exactly non-dimensional distance of 0.6 from left wall. Figure (\ref{valid_collimated}) shows the contour of irradiation which clearly shows the travel of collimated without any attenuation. For second step, fluid flow with heat transfer (without any radiation) is validated against Aswatha et al. \cite{Aswatha} and combined diffuse radiation and natural convection in a cavity whose top and bottom walls are adiabatic and vertical walls are isothermal at differential temperatures and radiatively opaque has been validated against Lari et al. \cite{Lari}. The present results for both the cases (see Fig. \ref{valid_temp}) are in good agreement with the published results.

\section{Grid independent test}

Numerical solutions of Navier-Stokes and energy equation and radiation transfer equation are sensitive to the spatial discretization. Additionally, radiative transfer equation also requires angular space discretization which provides directions along which radiation transfer equation is being solved. Thus, optimum number of grids and directions have been obtained through independent test study in two steps,
\begin{enumerate}
    \item Spatial grids independence test:
    Three spatial grid sizes,i.e, 60$\times$60, 80$\times$80 and 100$\times$100 are chosen to calculate the average Nusselt number on the bottom wall for the present problem of natural convection. The Nusselt number values calculated for above three grids arrangement on the bottom wall are 6.544 6.634 and 6.659, respectively. The percentage error between the first and second is 1.35$\%$, whereas between second and third is 0.37$\%$. Thus, the spatial grid points i.e 80$\times$80 is selected for further study.
    \item Angular direction independence test: The polar discretization has no effect on the two-dimensional cases, thus OpenFOAM fixes the number of polar directions to 2, in one hemisphere of angular space, while azimuthal considered directions are 3, 5, and 7 for the angular direction independent were studied. The Nusselt numbers on 80$\times$80 spatial grid and 2$\times$3, 2$\times$5, 2$\times$7 are 6.946, 7.0 and 7.012, respectively. The percentage difference in the first and second angular discretization is 0.7$\%$, whereas in second and third angular discreitization is 0.17$\%$. Thus, finally $n_\theta \times n_\phi =2 \times 5$ in one hemisphere angular space is selected. 
\end{enumerate}

\section{Results and Discussion}

Two scenarios of boundary conditions on semitransparent wall that are discussed in the section 2, have been simulated with collimated beam feature developed in OpenFOAM \cite{openfoam2017open} framework with natural convection and comprehensive results have been presented in this section.  

\subsection{Irradiation contour}

\begin{figure}[!t]
    \centering\includegraphics[width=7cm,scale=1]{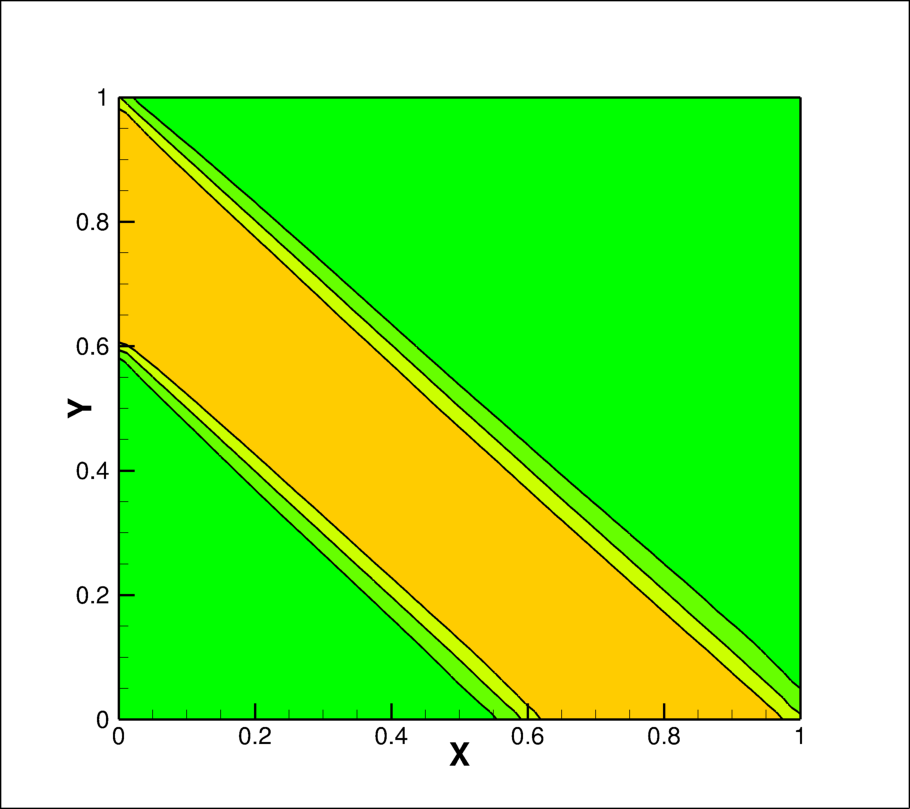}
    \label{G_C_N1000}
  \caption{Contours of collimated irradiation on semitransparent window at angle of $45^0$}
\label{G_qc_collimated}
\end{figure} 

A collimated beam in the direction of $45^0$ is applied on the semitransparent window and allowed to enter into the domain. The collimated beam in the cavity has been depicted by irradiation contours in Fig. \ref{G_qc_collimated}. Being transparent medium inside the cavity, the collimated beam strikes on the bottom wall and irradiation value remains constant through out the travel of the beam. The variations of irradiation value are seen at the edge of the irradiation contours that is basically due to rendering of the contours otherwise, contour change in the travel zone and outside should be like heavy side step function. The beam strikes at the non-dimensional distance 0.6 from the left corner and strike length of the collimated beam is 0.4 on the bottom wall which are same as window non-dimensional height and width. The effect of different values of irradiation and boundary conditions on semitransparent wall on the fluid flow and heat transfer characteristics will be explained in the subsequent sections.

\subsection{Case A: Conductively Adiabatic Condition}
The semitransparent wall behaves as only conductively adiabatic and allows the energy leaves by radiation mode. The detailed analysis of fluid flow and heat transfer are presented in the following section.

\subsubsection{Fluid Flow Characteristics}

\begin{figure}[!t]
 \begin{subfigure}{5cm}
    \centering\includegraphics[width=5cm]{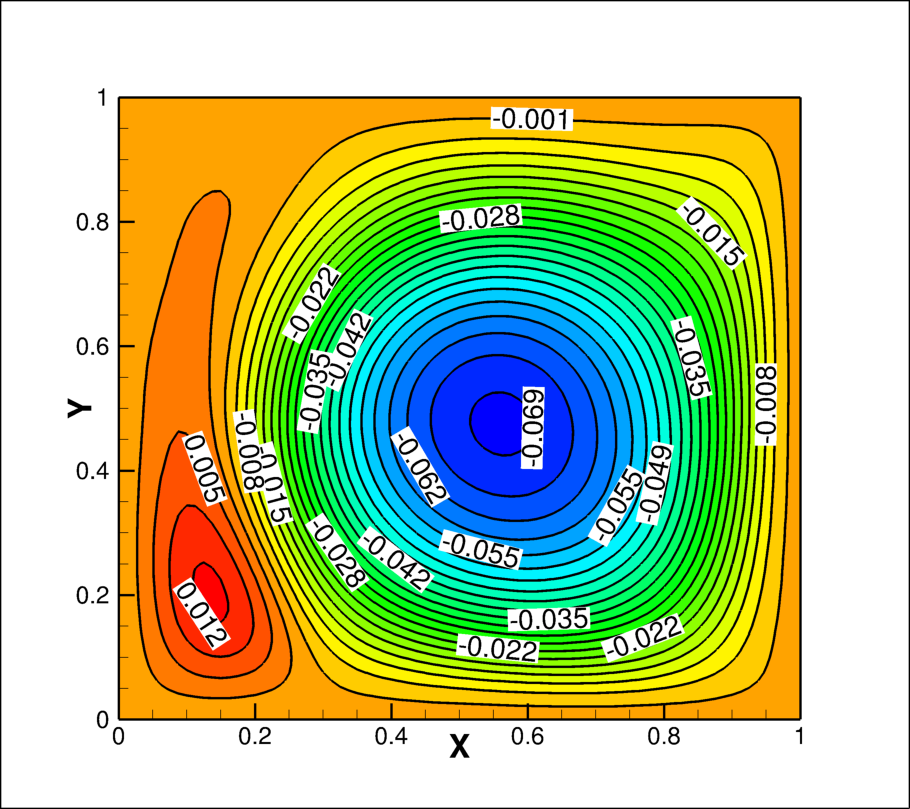}
    \caption{}
    \label{SF_C_N0}
  \end{subfigure}
   \begin{subfigure}{7cm}
    \centering\includegraphics[width=5cm]{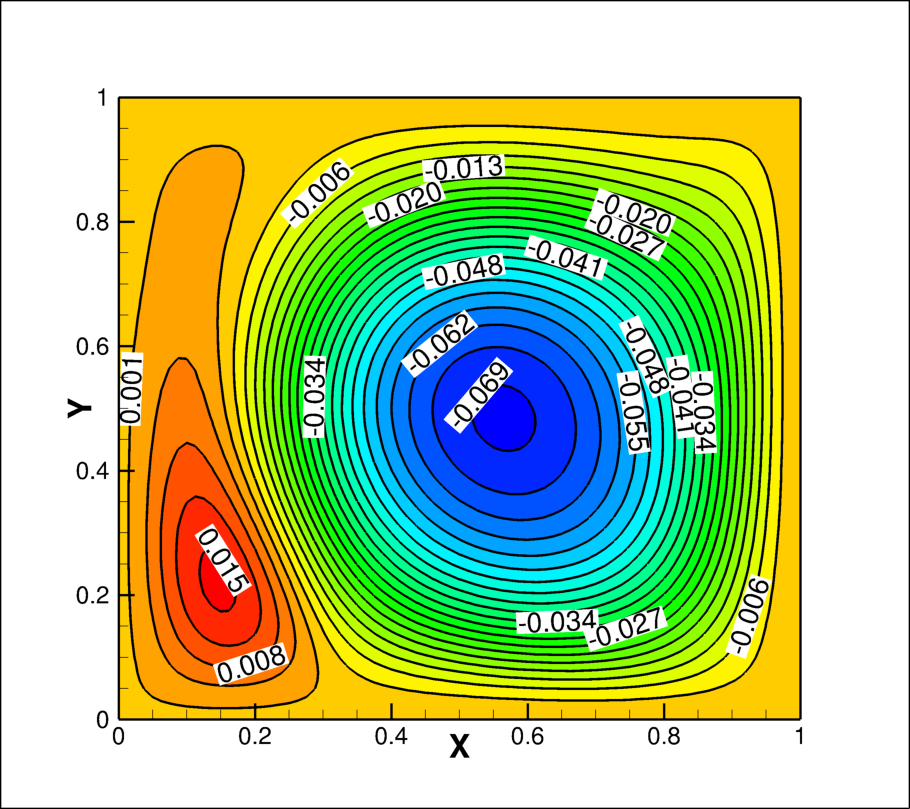}
    \caption{}
    \label{SF_C_N100}
  \end{subfigure}
  \begin{subfigure}{5cm}
    \centering\includegraphics[width=5cm]{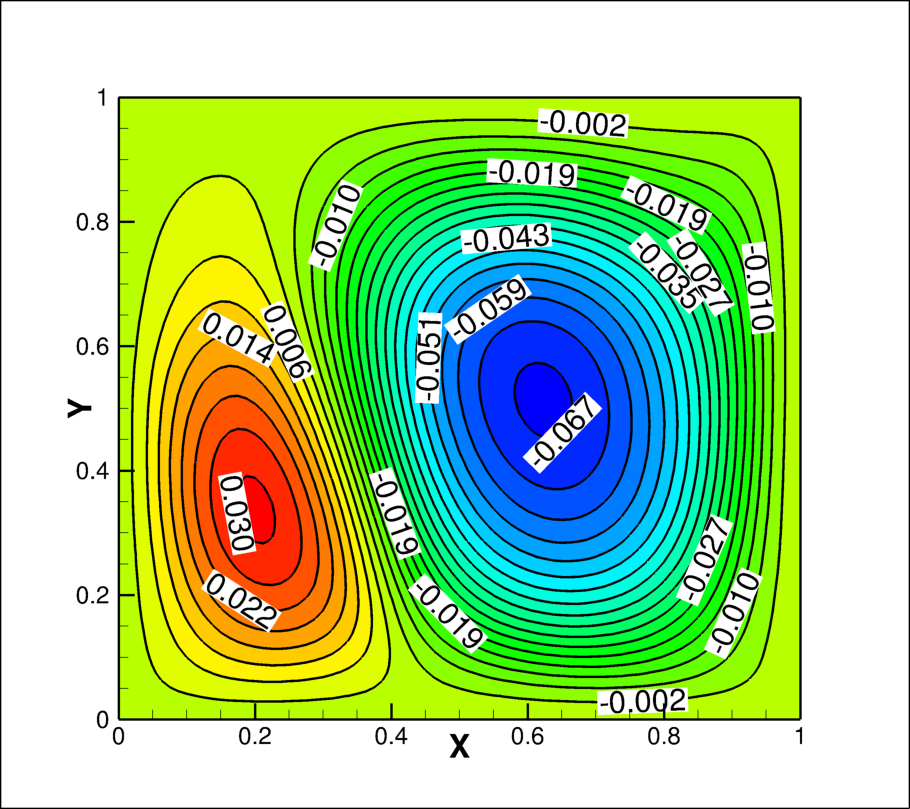}
    \caption{}
    \label{SF_C_N500}
  \end{subfigure}
   \begin{subfigure}{7cm}
    \centering\includegraphics[width=5cm]{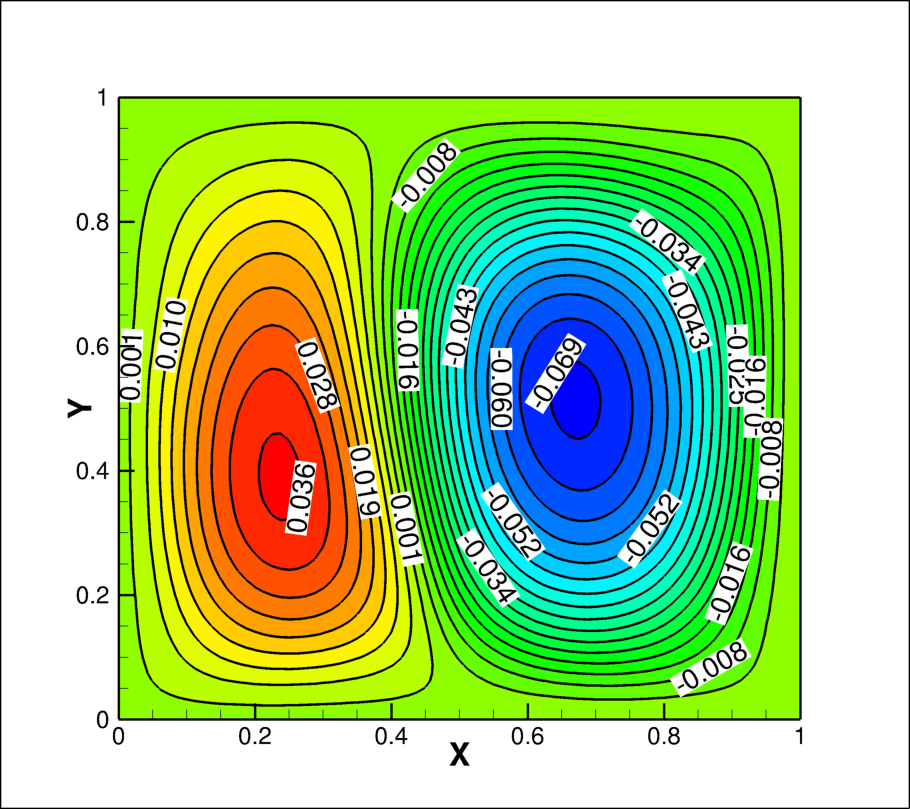}
    \caption{}
    \label{SF_C_N1000}
  \end{subfigure}
  \caption{Contours of non-dimensional stream function for collimated irradiation on the semitransparent window for the values of (a)G=0 (b)G=100 (c)G=500 and (d)G=1000 $W/m^2$}
\label{SF_qc_collimated}
\end{figure} 

The present problem consists two symmetrical counter rotating vortices inside the cavity when there would not have been collimated beam and the two vertical walls are isothermal and non-uniform heating at bottom \cite{Basak}. However, asymmetricity has been introduced by two ways (1) semitransparent vertical wall have been made conductively adiabatic and (2) there is collimated irradiation on the semitransparent window.  The asymmetric is high where size of left vortex is very small compared to right side vortex without any irradiation as shown in Fig. \ref{SF_C_N0}. This asymmeticity decreases with increase of irradiation values of 100, 500 as depicted in \ref{SF_C_N100}, \ref{SF_C_N500} \ref{SF_C_N1000}, respectively. The left vortex increases with the increase of irradiation value and the size of right vortex decreases in the similar proportion. This could be owing to the fact that being medium transparent, whole irradiation strikes on the bottom wall, this causes increase of the temperature of the bottom wall, thus, increase in the buoyancy force. The net of buoyancy and momentum force is now more in upward direction, therefore, the size of right vortex decreases. This net force is more in upward direction with the increase of the irradiation value. The maximum non-dimensional values of stream function for both the vortices for various values of irradiation are depicted in Table 1. The maximum non-dimensional value of stream function for right vortex remains almost constant while it increases for the left vortex. The flow rate in the left vortex increases with increase in the value of stream function and size of the left vortex while it decreases in the right vortex.

\begin{figure}[!t]
    \centering\includegraphics[width=6cm]{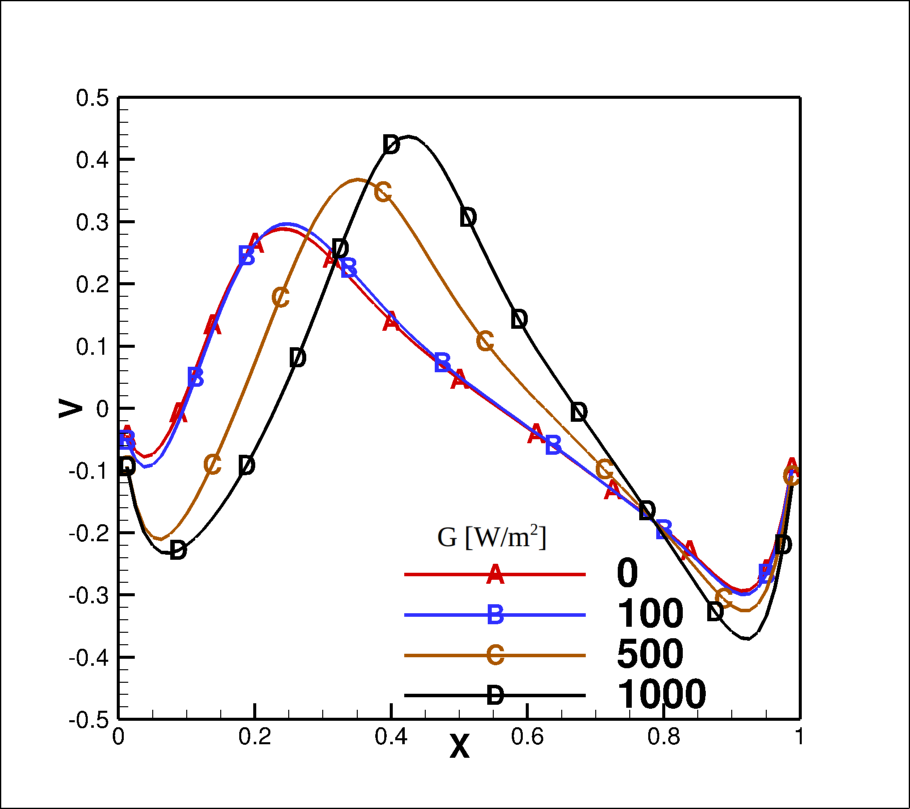}
    \caption{Variation of non-dimensional vertical velocity along the horizontal lines at the mid-height of the cavity for various values of irradiation}
    \label{qc_vertV_mid}
 \label{Mid_height_qc}
\end{figure} 

\begin{table}[!b]
\centering
\caption{Non-dimensional maximum stream function values for the various values of collimated irradiation}
\label{qc_SF_table}
\begin{tabular}{|ccccc|}
\hline
Irradiation(G) & 0 & 100 & 500 & 1000 \\ \hline
Right vortex & -0.069 & -0.069 & -0.067 & -0.069 \\ \hline
Left vortex & 0.012 & 0.015 & 0.03 & 0.036 \\ \hline
\end{tabular}
\end{table}

Figure \ref{qc_vertV_mid} depicts the variation of non-dimensional vertical velocity in the horizontal direction at the mid height of the cavity for various values of irradiation. The direction of vertical velocity is downward on both the vertical walls clearly indicate that hot fluid descends from these isothermal wall. Furthermore, the values of maximum vertical velocity in downward direction is lower near to left wall compared to right wall while maximum non-dimensional vertical velocity in downward direction increases with increase of the irradiation value. This increment is more near to left wall than the right wall, also the location of maximum downward vertical velocity remain same from both the vertical walls, i.e., around non-dimensional distance of 0.15 from both the vertical walls. The maximum non-dimensional vertical velocity values in downward direction near to left side wall are 0.08, 0.09, 0.2 and 0.22 for irradiation values of 0, 100, 500, 1000 $W/m^2$, respectively, while same values near to the right side walls are 0.3, 0.3, 0.32 and 0.38, respectively. There is also gain in maximum value of vertical velocity in upward direction, but unlike to the downward vertical velocity, the location of maximum upward velocity are different for different values of irradiation. These locations are at non-dimensional distance of 0.21, 0.21, 0.405, and 0.44, respectively from the left wall and maximum values of non-dimensional vertical  maximum velocity 0.3, 0.3, 0.35 and 0.42 for irradiation values of 0, 100, 500, 1000 $W/m^2$, respectively.

\subsubsection{Heat transfer characteristics}

\begin{figure}[!t]
\begin{subfigure}{5cm}
    \centering\includegraphics[width=5cm]{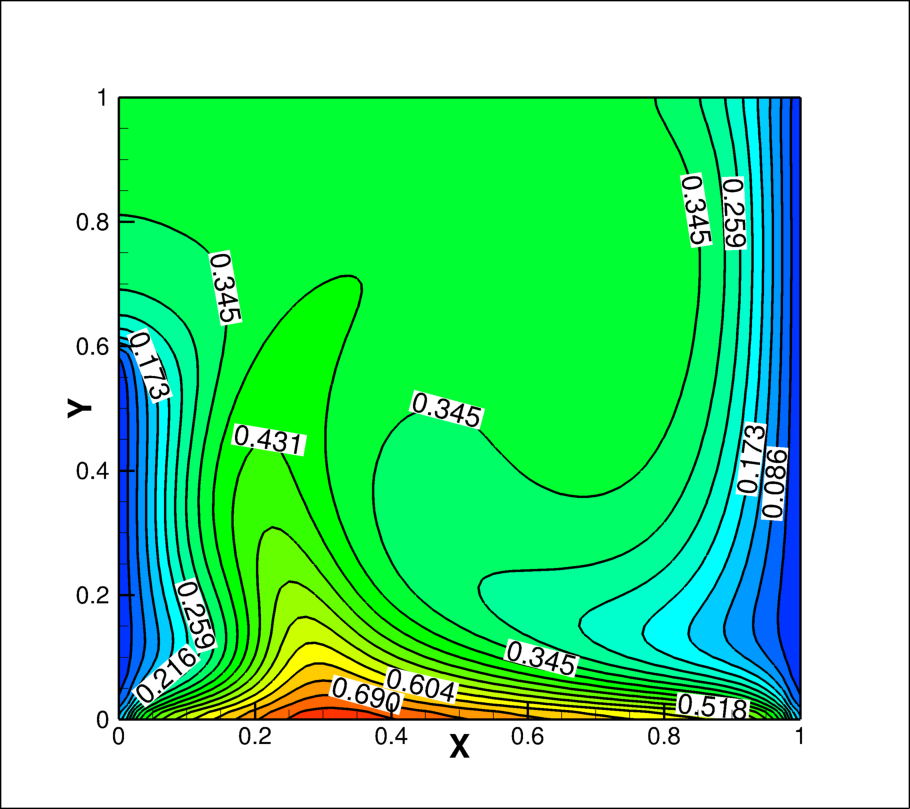}
    \caption{}
    \label{Temp_C_N0}
  \end{subfigure}
   \begin{subfigure}{8cm}
    \centering\includegraphics[width=5cm]{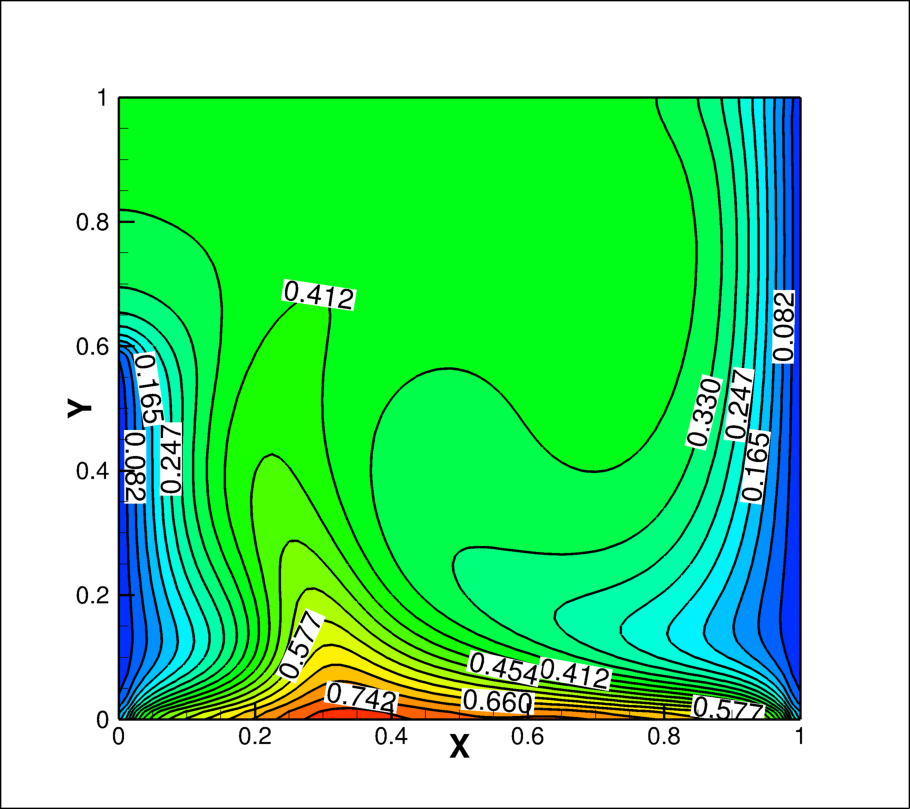}
    \caption{}
    \label{Temp_C_N100}
  \end{subfigure}
  \begin{subfigure}{5cm}
    \centering\includegraphics[width=5cm]{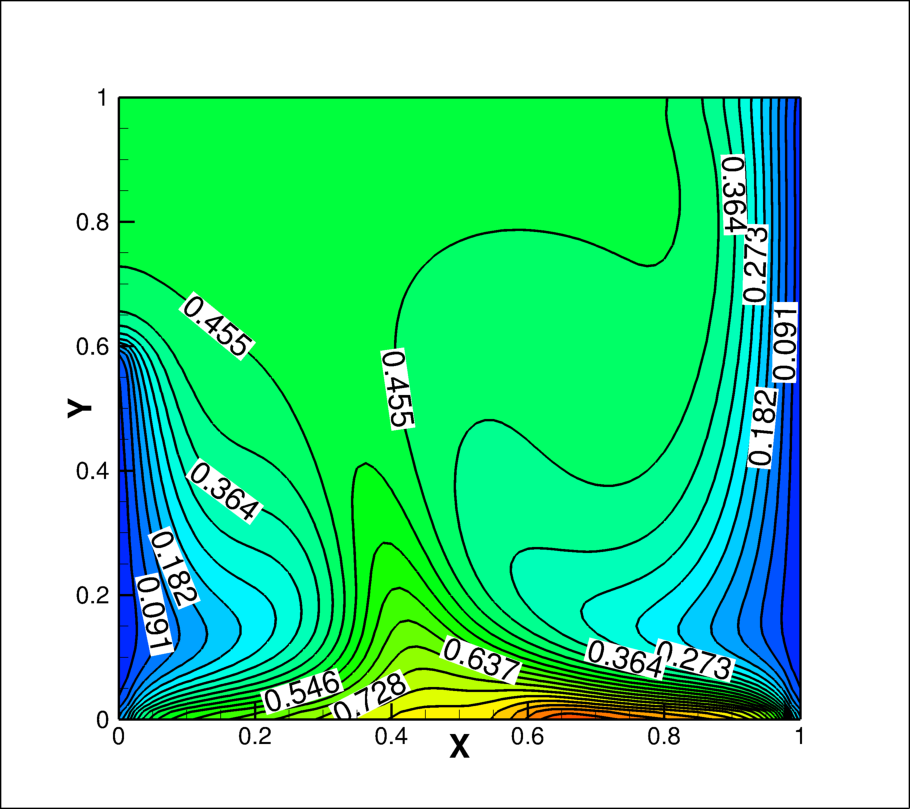}
    \caption{}
    \label{Temp_C_N500}
  \end{subfigure}
   \begin{subfigure}{8cm}
    \centering\includegraphics[width=5cm]{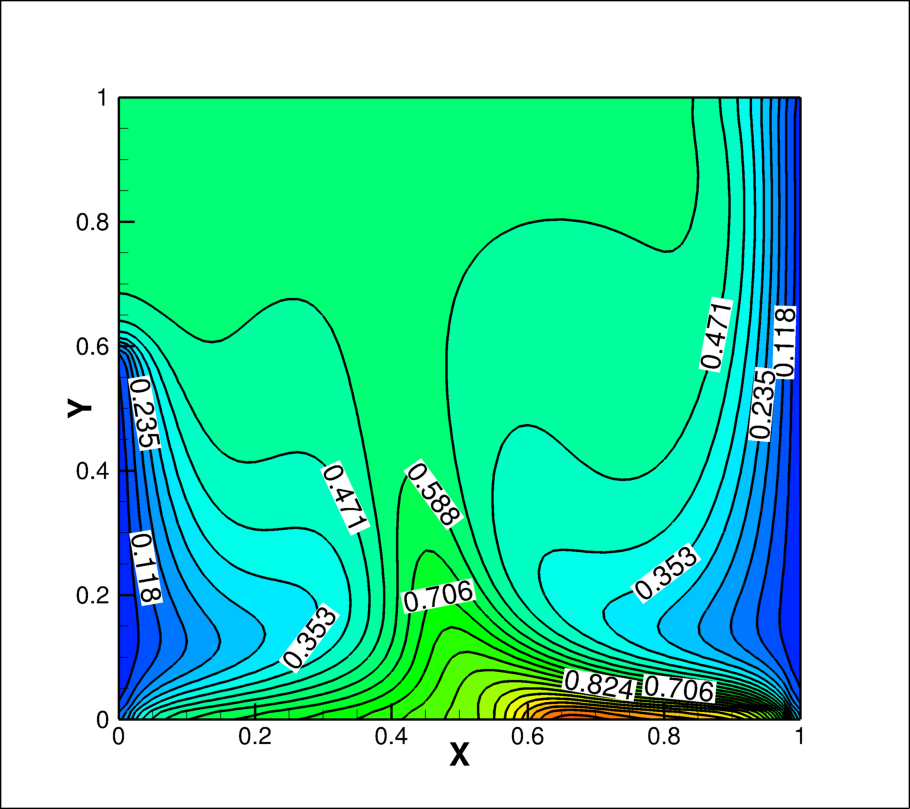}
    \caption{}
    \label{Temp_C_N1000}
  \end{subfigure}
  \caption{Contours of non-dimensional temperature for collimated irradiation on the semitransparent window of values (a)G=0 (b)G=100 (c)G=500 and (d)G=1000 $W/m^2$}
\label{Temp_qc_collimated}
\end{figure}

Being stream lines asymmetric (Fig. \ref{SF_qc_collimated}), the isothermal lines are also asymmetric (Fig. \ref{Temp_qc_collimated}). The isothermal lines at core are bent toward the left wall due to fact that the left vortex is smaller than the right vortex. Furthermore, the isotherms lines are clustered near to the bottom and isothermal walls whereas almost uniform temperature is spread in the core of the cavity near to adiabatic top wall. The isotherm lines are perpendicular to the semitransparent wall reveals that the wall is conductively adiabatic whereas the isotherms are not perpendicular to the upper wall, because upper wall is combinedly conductively and radiatively adiabatic. Table \ref{qc_Iso_table} represents the maximum non-dimensional temperature inside cavity and this increases with increase of irradiation value. This maximum non-dimensional temperature exists at the junction point of two vortices that creates a hot spot in the area of irradiation values of 0 and 100 $W/m^2$ whereas this maximum non-dimensional temperature exist at the strike length of collimated beam on the bottom wall for case of irradiation values of 500 and 1000 $W/m^2$ cases. One interesting point to notice is that isotherm lines are more clustered at the strike length of the collimated beam and also the non-dimensional temperature increases beyond to 1 for irradiation values of 500 to 1000 $W/m^2$.

\begin{figure}[!t]
 \begin{subfigure}{5cm}
    \centering\includegraphics[width=5cm]{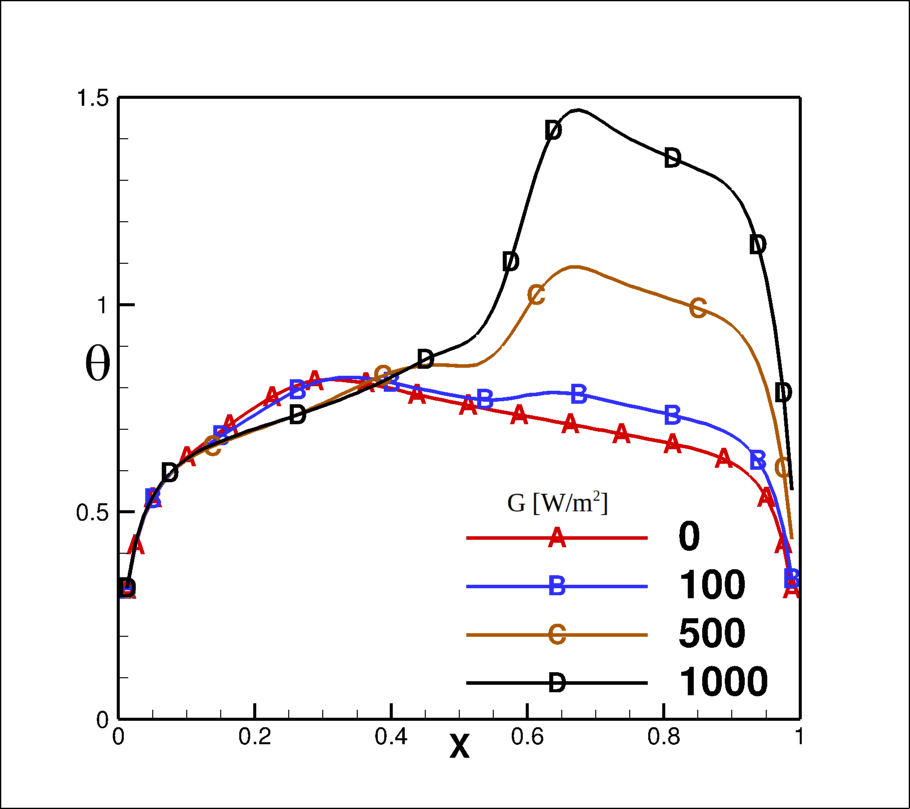}
    \caption{}
    \label{qc_bot_temp}
  \end{subfigure}
  \begin{subfigure}{7cm}
    \centering\includegraphics[width=5cm]{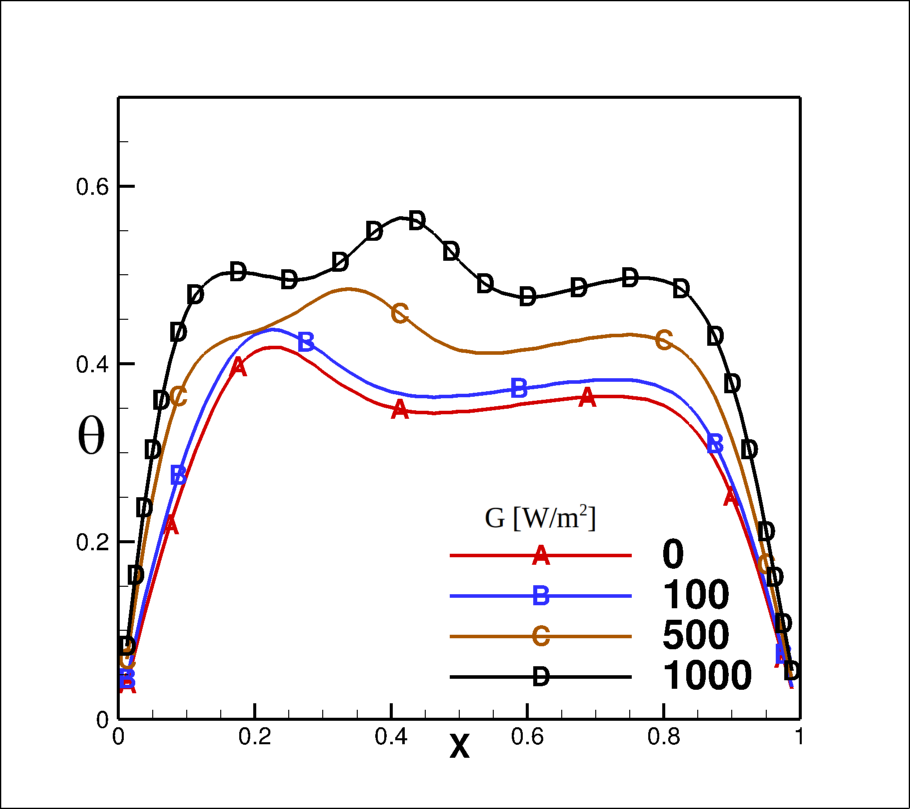}
    \caption{}
    \label{qc_mid_temp}
  \end{subfigure}
   \begin{subfigure}{5cm}
    \centering\includegraphics[width=5cm]{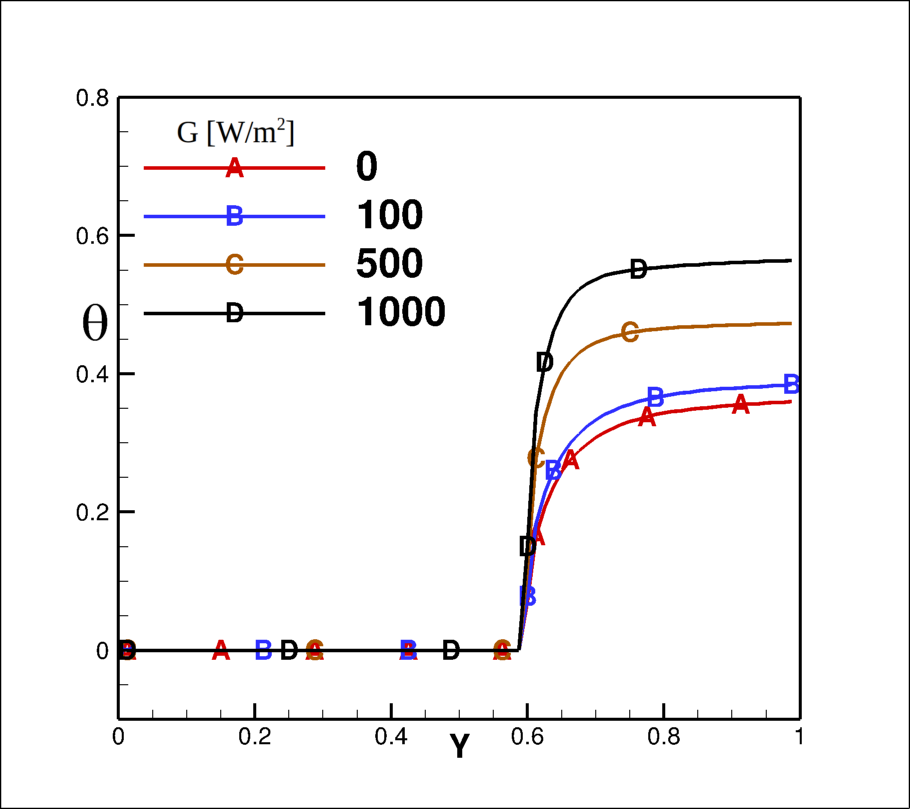}
    \caption{}
    \label{qc_left_temp}
  \end{subfigure}
  \begin{subfigure}{7cm}
    \centering\includegraphics[width=5cm]{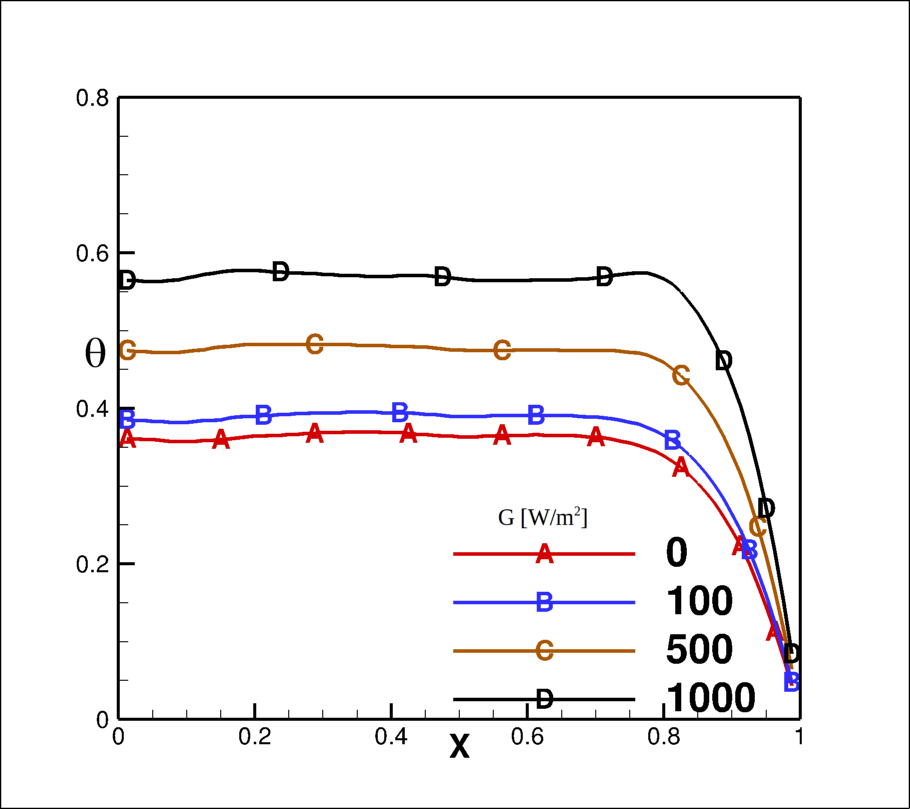}
    \caption{}
    \label{qc_top_temp}
  \end{subfigure}
  \caption{Variation of non-dimensional temperature at (a) bottom wall (b) horizontal line at mid-height of cavity (c) left wall and (d) top wall for various values of collimated irradiation}
\label{qc_temp_walls}
\end{figure} 

\begin{table}[!b]
\centering
\caption{Non-dimensional maximum isothermal values for the various values of collimated irradiation}
\label{qc_Iso_table}
\begin{tabular}{|ccccc|}
\hline
Irradiation(G) & 0 & 100 & 500 & 1000 \\ \hline
Maximum & 0.820 & 0.830 & 1.095 & 1.426 \\ \hline
\end{tabular}
\end{table}

The variation of non-dimensional temperature on the various walls and also along the horizontal direction at the mid height of the cavity have been shown in Fig \ref{qc_temp_walls}. The non-dimensional temperature at the bottom wall increases from the left corner and reaches to peak value of 0.82 then decreases upto the right corner for case of $G=0$ while there is small rise in temperature at the strike length of collimated beam for G=100 $W/m^2$. Whereas there is drastic rise in non-dimensional temperature at the strike length of collimated beam for irradiation value of 500 $W/m^2$ and 1000 $W/m^2$ and maximum non-dimensional temperature reaches to 1.2 and 1.47, respectively (see Fig. \ref{qc_bot_temp}) at the bottom wall. The temperature rise is not that much drastic in the path of collimated beam at the mid height of the cavity (Fig. \ref{qc_mid_temp}). The global maxima in the temperature curve is near to left wall at mid-height of the cavity that shifts towards right with the increase of the irradiation and the maximum non-dimensional temperatures are at 0.4, 0.42, 5.2 and 5.6 at non-dimensional distance from left is 0.2, 0.2, 0.38, 0.42 for irradiation values of 0, 100, 500 and 1000 $W/m^2$, respectively. The non-dimensional temperature remain zero on the left wall upto the height of 0.6 because of isothermal condition, afterwards the non-dimensional temperature increases all of sudden (Fig. \ref{qc_left_temp}). The maximum non-dimensional temperature rise are 0.38, 0.39, 0.41 and 0.5 for irradiation values of 0, 100 , 500 and 1000 $W/m^2$, respectively and remain more or less constant afterwards on the left semitransparent wall. The non-dimensional temperature remains constant and matches with temperature rise on semitransparent vertical wall for most of the length on upper wall, only decrease towards end to match the temperature of the isothermal right wall (see Fig. \ref{qc_top_temp})

\subsubsection{Nusselt number}

\begin{figure}[!b]
 \begin{subfigure}{5cm}
    \centering\includegraphics[width=5cm]{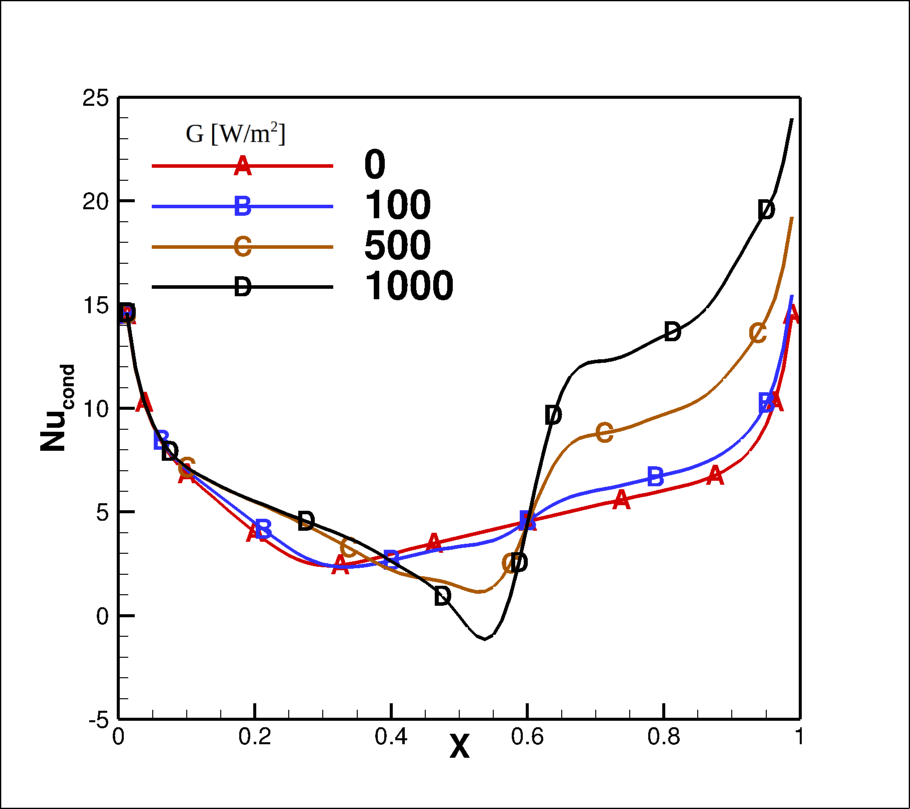}
    \caption{}
    \label{qc_bot_cond}
  \end{subfigure}
   \begin{subfigure}{5cm}
    \centering\includegraphics[width=5cm]{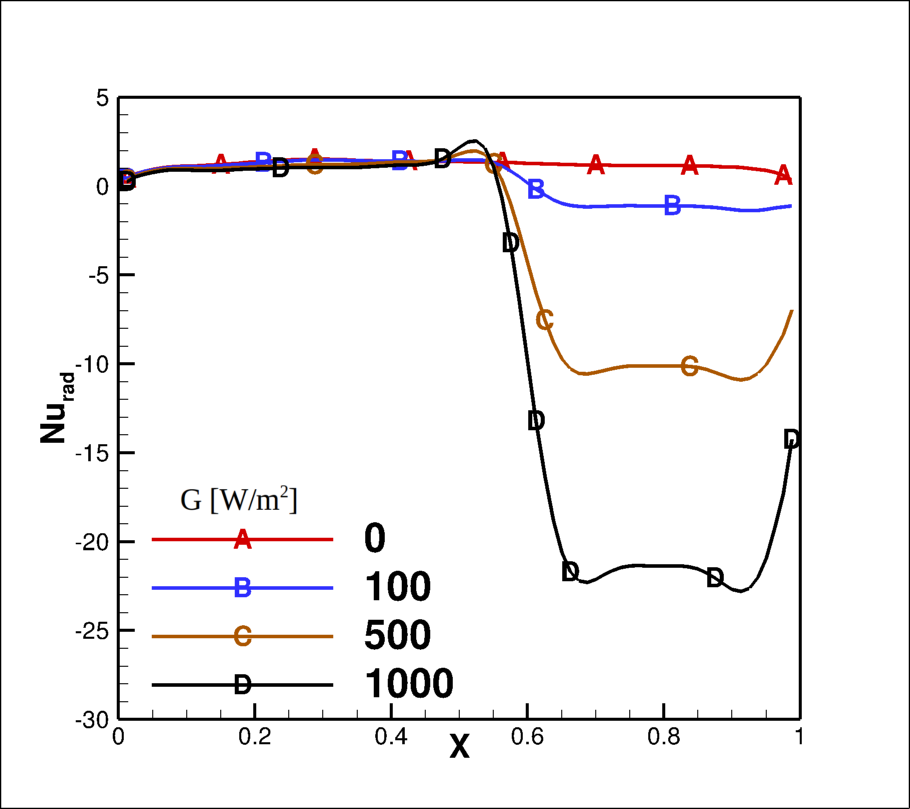}
    \caption{}
    \label{qc_bot_rad}
  \end{subfigure}
  \begin{subfigure}{12cm}
    \centering\includegraphics[width=5cm]{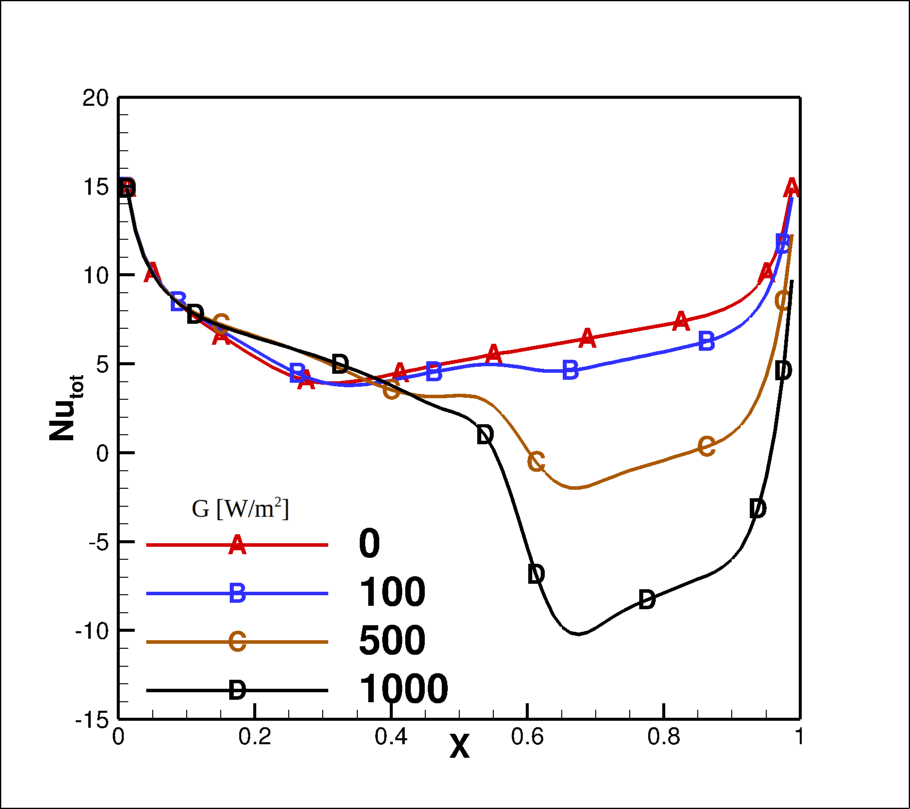}
    \caption{}
    \label{qc_bot_tot}
  \end{subfigure}
  \caption{Variation of (a) conduction (b) radiation and (c) total Nusselt number on the bottom wall for various values of collimated irradiation on semitransparent window}
\label{qc_bot_Nu}
\end{figure} 

\begin{figure}[!htb]
 \begin{subfigure}{5cm}
    \centering\includegraphics[width=5cm]{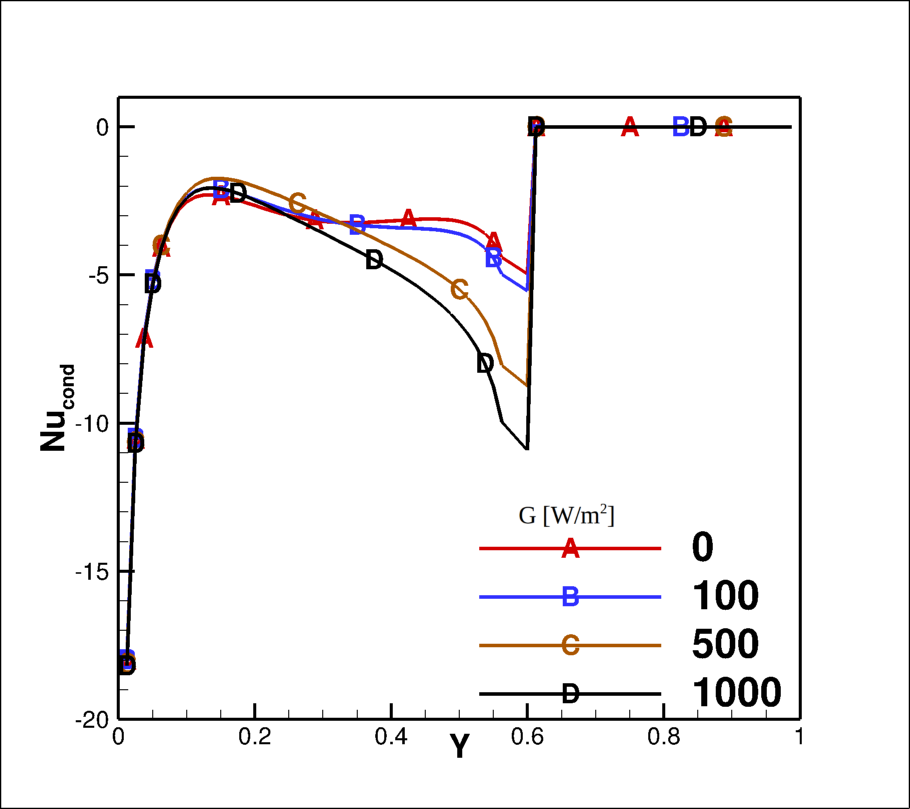}
    \caption{}
    \label{qc_left_cond}
  \end{subfigure}
   \begin{subfigure}{5cm}
    \centering\includegraphics[width=5cm]{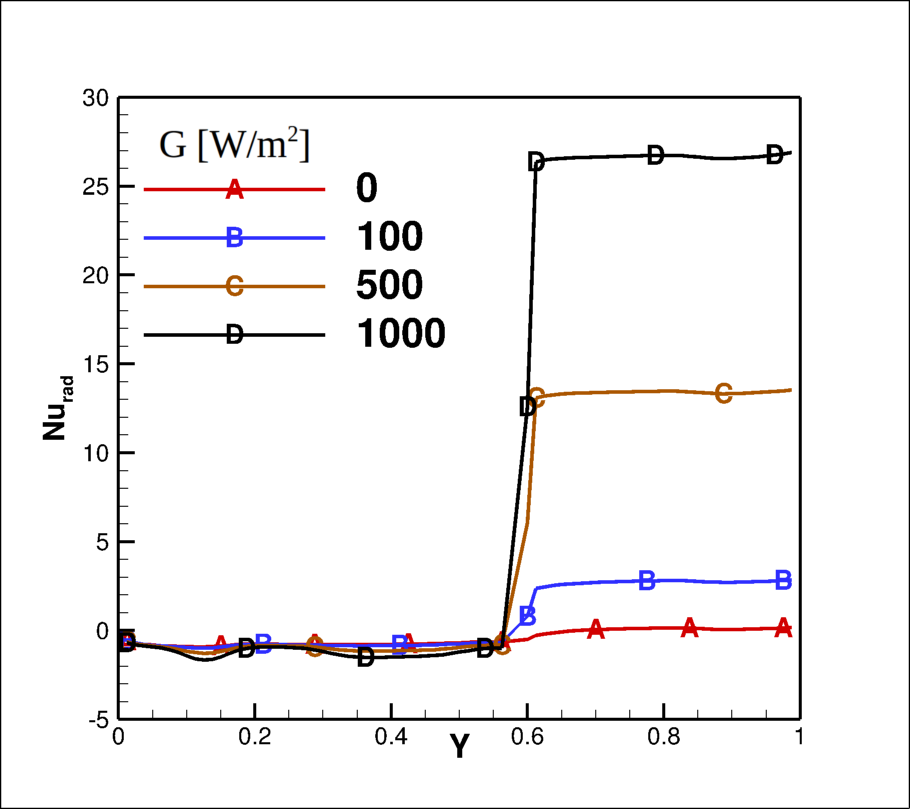}
    \caption{}
    \label{qc_left_rad}
  \end{subfigure}
  \begin{subfigure}{12cm}
    \centering\includegraphics[width=5cm]{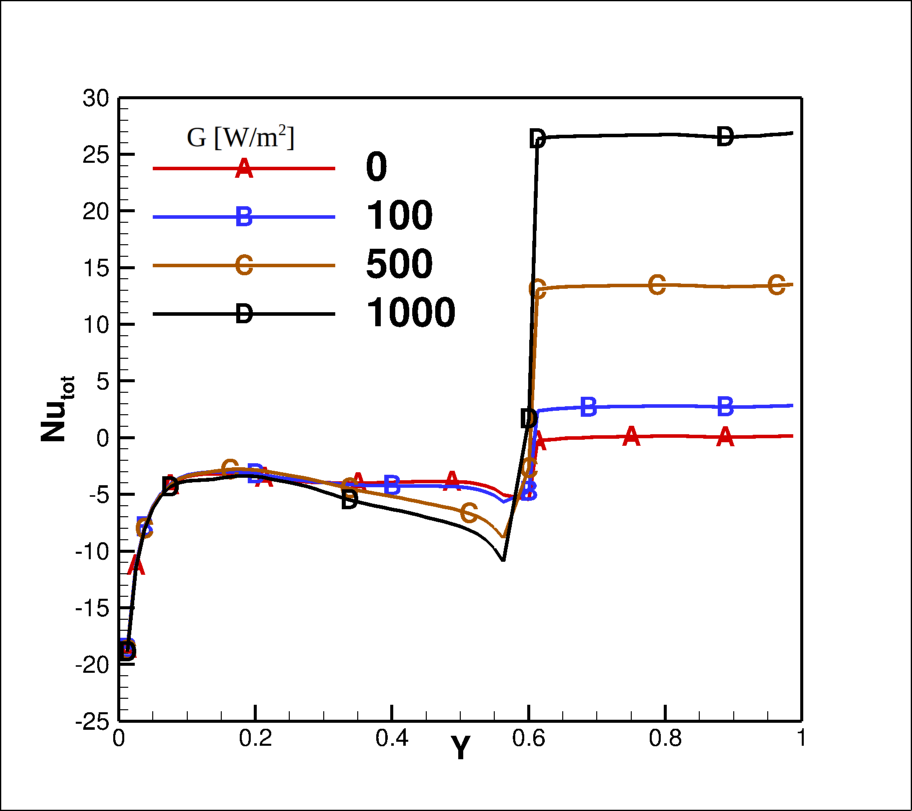}
    \caption{}
    \label{qc_left_tot}
  \end{subfigure}
  \caption{Variation of (a) conduction (b) radiation and (c) total Nusselt number on the left wall for various values of collimated irradiation on semitransparent window}
\label{qc_left_Nu}
\end{figure}

The variations of conduction, radiation and total Nusselt numbers on the bottom wall are shown in Fig. \ref{qc_bot_Nu}(a), (b) and (c), respectively, for various values of irradiation. The conduction Nusselt number is very high at the end of the walls and monotonically decreases from both the ends and reached to minimum values at different locations on the wall for different values of irradiation; its location is at the junction of two vortices for irradiation value of 0 and 100 $W/m^2$ and before the strike point from left for irradiation values of 500 and 1000 $W/m^2$ afterward there is sudden rise in the value of conduction Nusselt numbers. The minimum conduction Nusselt numbers are same for the irradiation values of 0 and 100 $W/m^2$ while for irradiation values of 500 and 1000 $W/m^2$ its value are 2 and 0 respectively. While maximum conduction Nusselt numbers are also same for irradiation value of 0 and 100 $W/m^2$ which is 15 while its value is 20 and 24 for irradiation value of 500 and 1000 $W/m^2$, respectively, this maximum Nusselt number is found near to right corner of the wall. The radiative Nusselt number is almost zero upto the strike point from left corner of wall and sudden increase in the radiation Nusselt number has been observed over the strike length of the collimated beam on the bottom wall. The negative radiative Nusselt number indicates that energy is going out of the domain by radiation mode of heat transfer. The increase of radiative Nusselt number happens with increase in irradiation value. Also, the radiative Nusselt number is zero over whole curve length of the bottom wall for irradiation value  0 $W/m^2$ indicates the contribution of the diffuse radiation is almost negligible in the present problem. Thus, the total Nusselt number is governed by conduction Nusselt number upto the non strike length of the collimated afterword, it is governed by radiation over the strike length of beam, if the irradiation values is sufficiently high, here this happens for $G > 500$ $W/m^2$ (Fig. \ref{qc_bot_tot}).  Similarly, the conduction, radiation and total Nusselt number on the left wall are shown in Fig. \ref{qc_left_Nu}(a), (b) and (c), respectively. The conduction Nusselt number decreases drastically upto non-dimensional height of 0.1 afterward its starts increasing slowly for irradiation values of 0 and 100 $W/m^2$ while fast increment happens for higher irradiation values of 500 and 1000 $W/m^2$ upto the lower isothermal wall height, afterwards, sudden decrease to zero due to conductively adiabatic wall condition on the upper semitransparent wall. The negative conduction Nusselt number indicate that energy goes out through conduction from lower isothermal wall. The radiative Nusselt number is almost zero on the isothermal wall further emphasises that contribution of diffuse radiation is almost zero. While there is sudden increase of radiative Nusselt number on the upper semitransparent wall and remain constant over the height due to collimated irradiation. The positive radiative Nusselt number indicates that energy enters into the domain through radiative mode of heat transfer. The total Nusselt number is linear sum of both conduction and radiation Nusselt numbers, thus, conduction is dominant on lower isothermal wall and radiation on the upper semitransparent wall. Thus, net energy leaves out through isothermal wall and comes into the domain through upper semitransparent wall.

\begin{figure}[!htb]
 \begin{subfigure}{5cm}
    \centering\includegraphics[width=5cm]{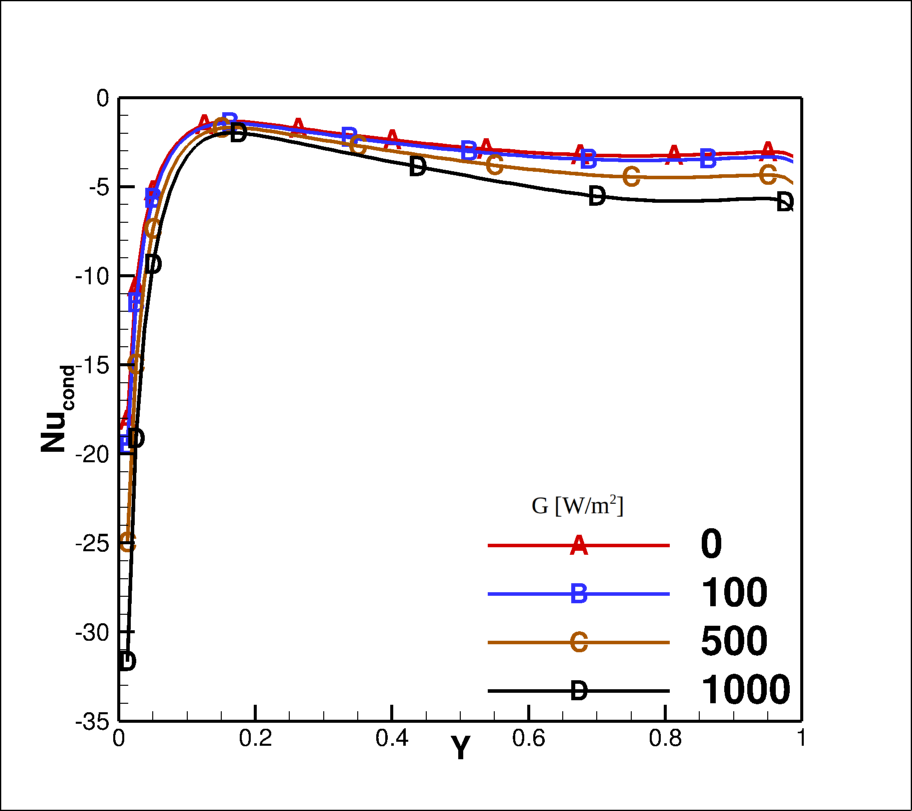}
    \caption{}
    \label{qc_right_cond}
  \end{subfigure}
   \begin{subfigure}{5cm}
    \centering\includegraphics[width=5cm]{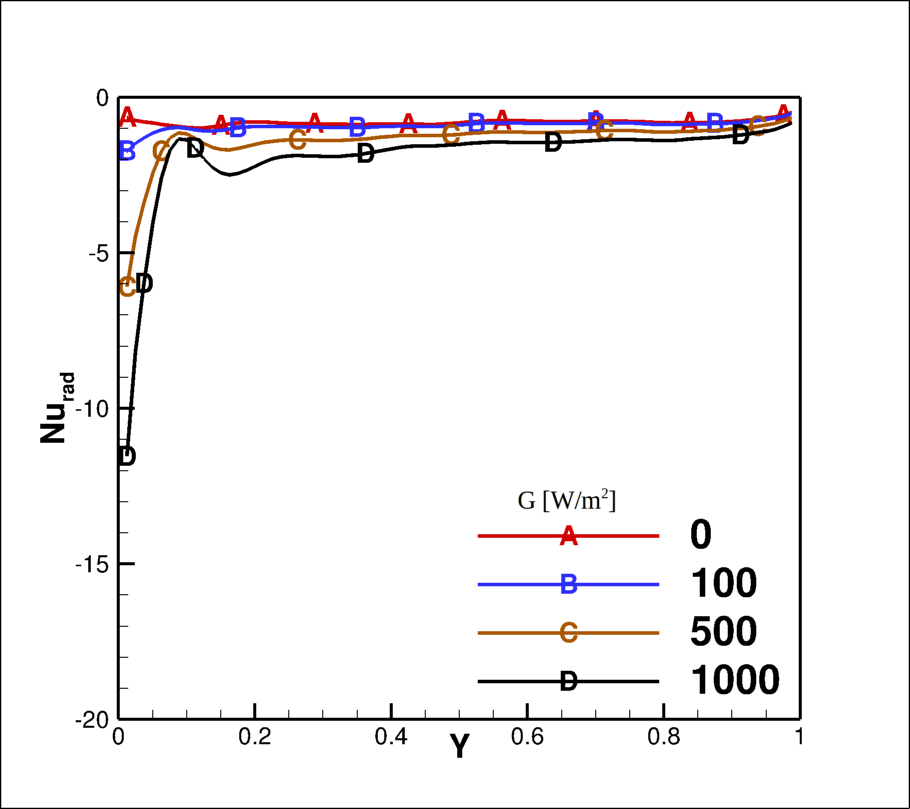}
    \caption{}
    \label{qc_right_rad}
  \end{subfigure}
  \begin{subfigure}{12cm}
    \centering\includegraphics[width=5cm]{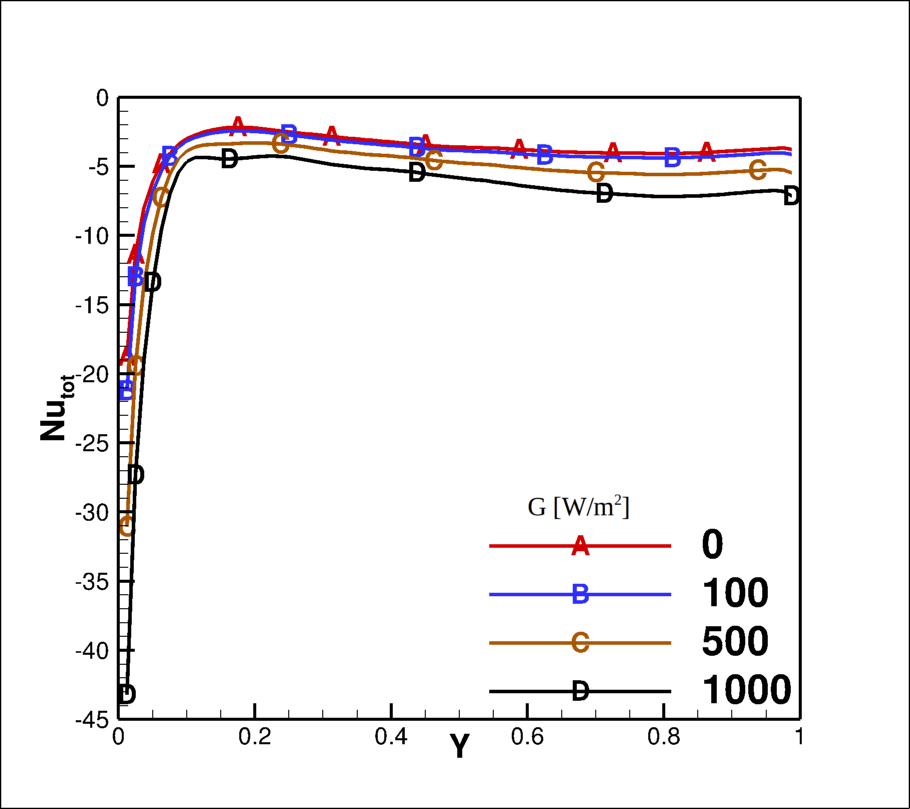}
    \caption{}
    \label{qc_right_tot}
  \end{subfigure}
  \caption{Variation of (a) conduction (b) radiation and (c) total Nusselt number on the right wall for various values of collimated irradiation on semitransparent window}
\label{qc_right_Nu}
\end{figure}

Further the conduction, radiation and total Nusselt number on right side wall are shown in Fig \ref{qc_right_Nu}(a), (b) and (c), respectively. The conduction Nusselt number decreases sharply to a value of 2 to a height of 0.1 afterwards, it almost remain constant throughout the height of wall for irradiation value of 100 $W/m^2$ and small increment is found for irradiation value of 500 and 1000 $W/m^2$. Whereas, the radiation Nusselt number remains almost constant for the irradiation values of 0 and 100 throughout the wall but large value of Nusselt number is found at the lower height of the wall then the value decreases to 2, within a non-dimensional height of 0.1 for the irradiation values of 500 and 1000 $W/m^2$, afterthat, the value almost remains constant and the Nusselt number value is very small. The total Nusselt number of the right wall is the linear combination of two Nusselt numbers i.e conduction and radiation Nusselt number. It can be clearly understood that variation of total Nusselt number is similar to conduction Nusselt number over whole height of right wall. The negative values show the energy leaves by both conduction and radiation mode of heat transfer from the right wall.

The average Nusselt number on the various walls of the cavity for different values of irradiation is shown in Table 3. The total average Nusselt number decreases with the increase of irradiation value on bottom wall while average conduction Nusselt number increases and radiation Nusselt number changes its sign from being positive on low irradiation value to negative for higher irradiation value. The left wall has two portions, lower part is isothermal where average of both conduction and radiation Nusselt numbers remain almost constant and negative, while upper portion being semitransparent, the  conduction Nusselt number is zero  due to conductively adiabatic condition and radiative Nusselt number increases with increase of irradiation value. Furthermore total average Nusselt number first being negative then becomes positive with increase of irradiation value on the left wall. The average conduction and radiation Nusselt numbers on the right wall increases with increase of irradiation values, thus the total Nusselt number also increases with negative sign. 

\begin{landscape}
\begin{table}[!t]
\centering
\caption{Average Nusselt number on different walls for various values of irradiation for conductively adiabatic semitransparent wall}
\label{qc_avg_Nu_table}
\begin{tabular}{|llllllllllll|}
\hline
\multicolumn{1}{|c|}{\multirow{3}{*}{\begin{tabular}[c]{@{}c@{}}Irradiation\\ (G)\end{tabular}}} & \multicolumn{3}{c|}{Bottom wall} & \multicolumn{5}{c|}{Left wall} & \multicolumn{3}{c|}{Right wall} \\ \cline{2-12} 
\multicolumn{1}{|c|}{} & \multicolumn{1}{c|}{\multirow{2}{*}{Conduction}} & \multicolumn{1}{c|}{\multirow{2}{*}{Radiation}} & \multicolumn{1}{c|}{\multirow{2}{*}{Total}} & \multicolumn{2}{c|}{Isothermal wall} & \multicolumn{2}{c|}{Semitransparent wall} & \multicolumn{1}{c|}{} & \multicolumn{1}{c|}{\multirow{2}{*}{Conduction}} & \multicolumn{1}{c|}{\multirow{2}{*}{Radiation}} & \multirow{2}{*}{Total} \\ \cline{5-9}
\multicolumn{1}{|c|}{} & \multicolumn{1}{c|}{} & \multicolumn{1}{c|}{} & \multicolumn{1}{c|}{} & \multicolumn{1}{c|}{Conduction} & \multicolumn{1}{c|}{Radiation} & \multicolumn{1}{c|}{Conduction} & \multicolumn{1}{c|}{Radiation} & \multicolumn{1}{c|}{Total} & \multicolumn{1}{c|}{} & \multicolumn{1}{c|}{} &  \\ \hline
0 & 5.403 & 0.923 & 6.326 & -2.138 & -0.435 & 0 & 0.026 & -2.547 & -2.975 & -0.791 & -3.766 \\ \hline
100 & 5.678 & 0.318 & 5.996 & -2.177 & -0.444 & 0 & 1.057 & -1.564 & -3.531 & -0.901 & -4.432 \\ \hline
500 & 6.792 & -3.435 & 3.357 & -2.702 & -0.478 & 0 & 5.188 & 2.008 & -3.991 & -1.335 & -5.326 \\ \hline
1000 & 8.15 & -7.98 & 0.17 & -2.852 & -0.521 & 0 & 10.326 & 6.953 & -5.244 & -1.879 & -7.123 \\ \hline
\end{tabular}
\end{table}
\end{landscape}

\subsection{Case B: Combinedly conductively and radiatively adiabatic condition}

In this section, the fluid flow and heat transfer characteristics have been extensively studied when combinedly conductively and radiatively adiabatic boundary condition has been applied on the semitransparent window. In this condition even energy does not leave the system by radiaitve mode of heat transfer through semitransparent window.

\subsubsection{Fluid flow characteristics}

\begin{figure}[!t]
 \begin{subfigure}{5cm}
    \centering\includegraphics[width=5cm]{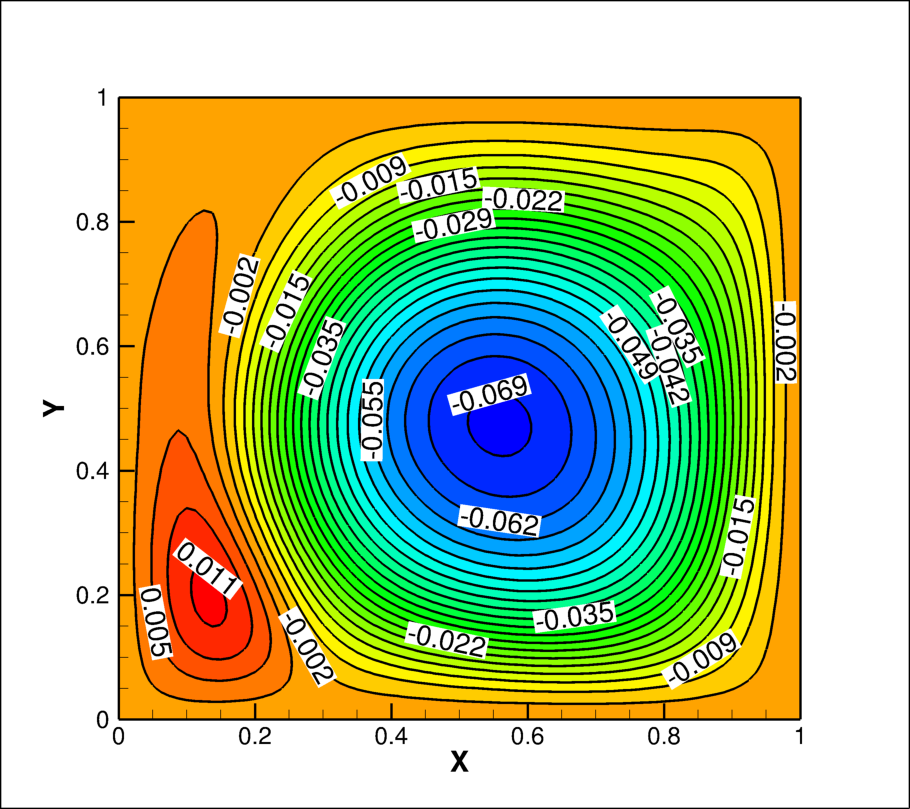}
    \caption{}
    \label{SF_C1_N0}
  \end{subfigure}
   \begin{subfigure}{7cm}
    \centering\includegraphics[width=5cm]{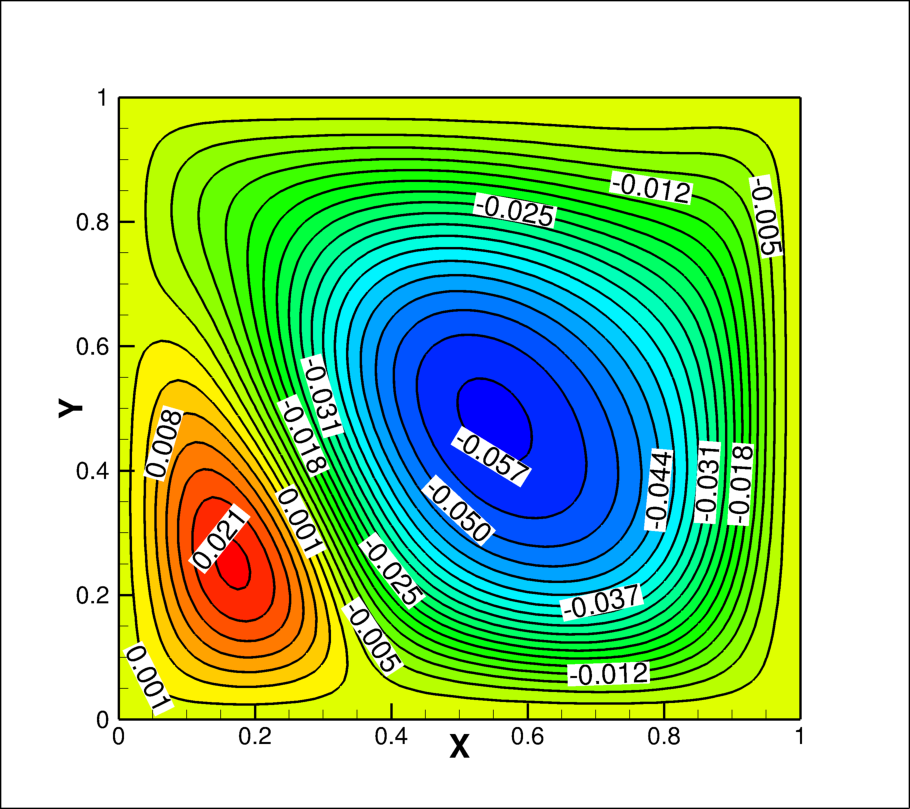}
    \caption{}
    \label{SF_C1_N100}
  \end{subfigure}
  \begin{subfigure}{5cm}
    \centering\includegraphics[width=5cm]{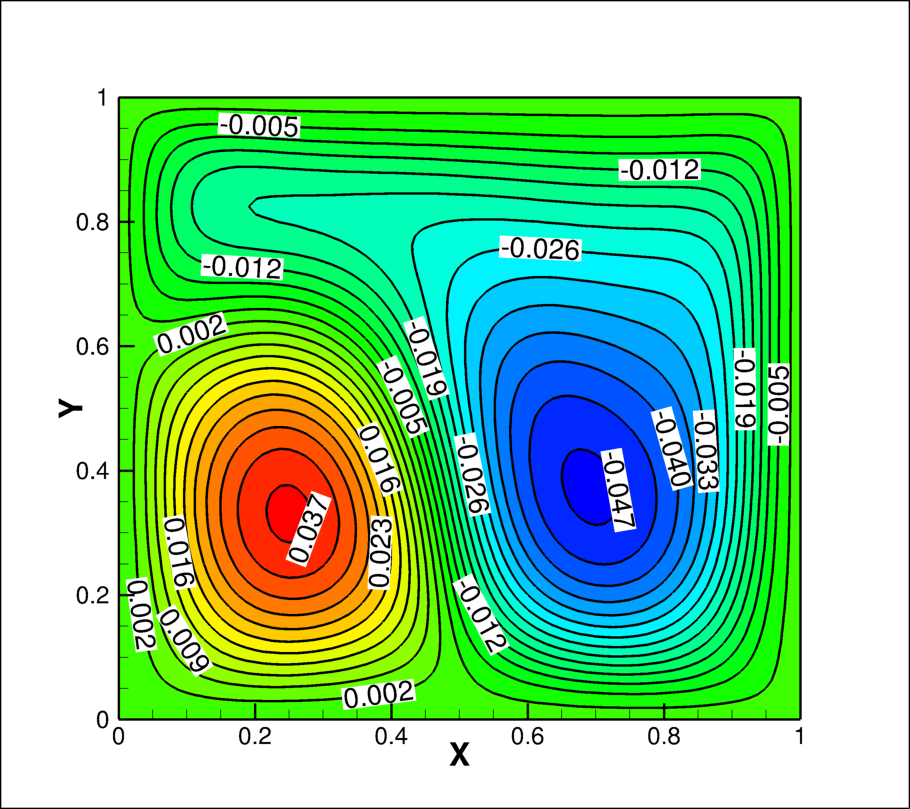}
    \caption{}
    \label{SF_C1_N500}
  \end{subfigure}
   \begin{subfigure}{7cm}
    \centering\includegraphics[width=5cm]{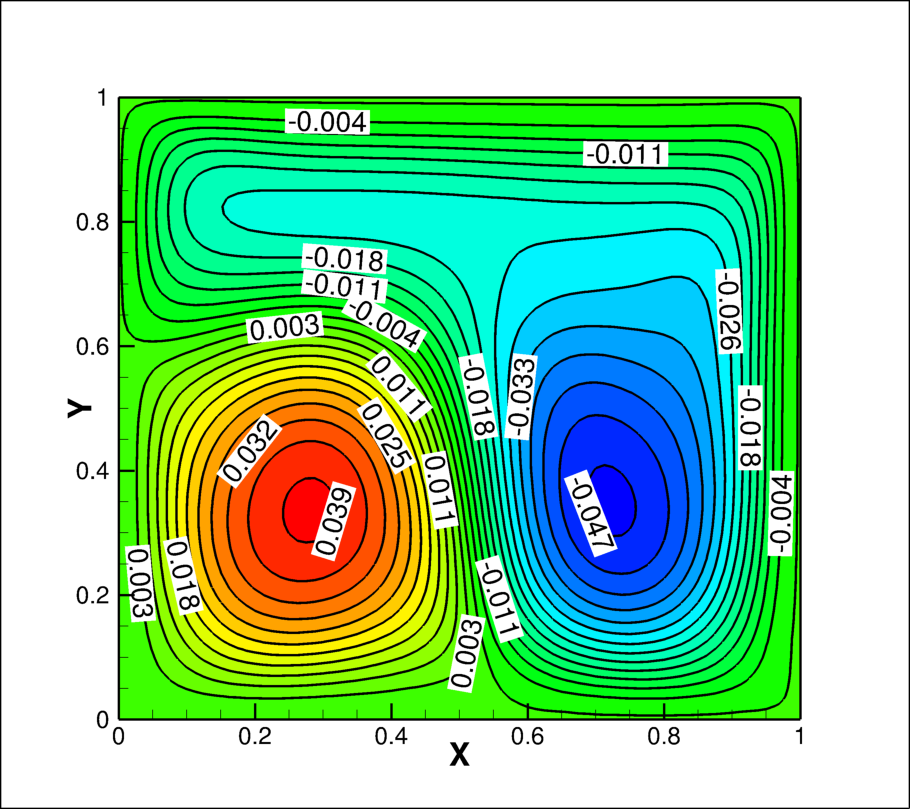}
    \caption{}
    \label{SF_C1_N1000}
  \end{subfigure}
  \caption{Contours of non-dimensional stream function for collimated irradiation of (a)G=0 (b)G=100 (c)G=500 and (d)G=1000 $W/m^2$ on semitransparent window}
\label{SF_qcqr_collimated}
\end{figure}

The contours of non-dimensional stream function for various values of irradiation on the semitransparent window have been presented in Fig. \ref{SF_qcqr_collimated}. The stream function contours for zero irradiation (Fig.\ref{SF_C1_N0}) is similar to case A (Fig. \ref{SF_C_N0}) because of negligible diffuse radiation is present in current problem. However, the behaviour of two vortices changes drastically with the collimated irradiation on the semitransparent window. The left side vortex always remains upto the height of isothermal wall for all values of irradiation, however, its width increases with the increase of irradiation values. Being combinedly adiabatic semitransparent wall, it is expected that it will be hotter, thus fluid rises from semitransparent wall and only lose energy to the right side of cold isothermal wall. Therefore, area occupied by right side vortex becomes quite large including whole area above to the lower isothermal wall and right side vortex area. At lower irradiation value $(G=100W/m^2)$ the orientation of right side vortex is diagonal, whereas the orientation of right side vortex becomes straight with increase of irradiation value and layered flow happen on the upper part (above to the lower left wall) of the cavity however, this layered flow is connected to the right side vortex. Table 4 shows the maximum value of non-dimensional stream function for these two vortices for various values of irradiation. The non-dimensional stream function value keeps on increasing for left vortex with increase of irradiation value while its decreases for right vortex with increase of irradiation value upto 500 $W/m^2$ then it remains constant. Furthermore, the area occupied by left vortex also increases which indicates that the volume flow rate increases in left vortex with increase of irradiation value.

\begin{table}[!t]
\centering
\caption{Non-dimensional maximum stream function values for the various values of collimated irradiation on semitransparent window}
\label{qcqr_SF_table}
\begin{tabular}{|ccccc|}
\hline
Irradiation(G) & 0 & 100 & 500 & 1000 \\ \hline
Right & -0.069 & -0.057 & -0.047 & -0.047 \\ \hline
Left & 0.011 & 0.021 & 0.037 & 0.039 \\ \hline
\end{tabular}
\end{table}

\begin{figure}[!t]  
    \centering\includegraphics[width=6cm]{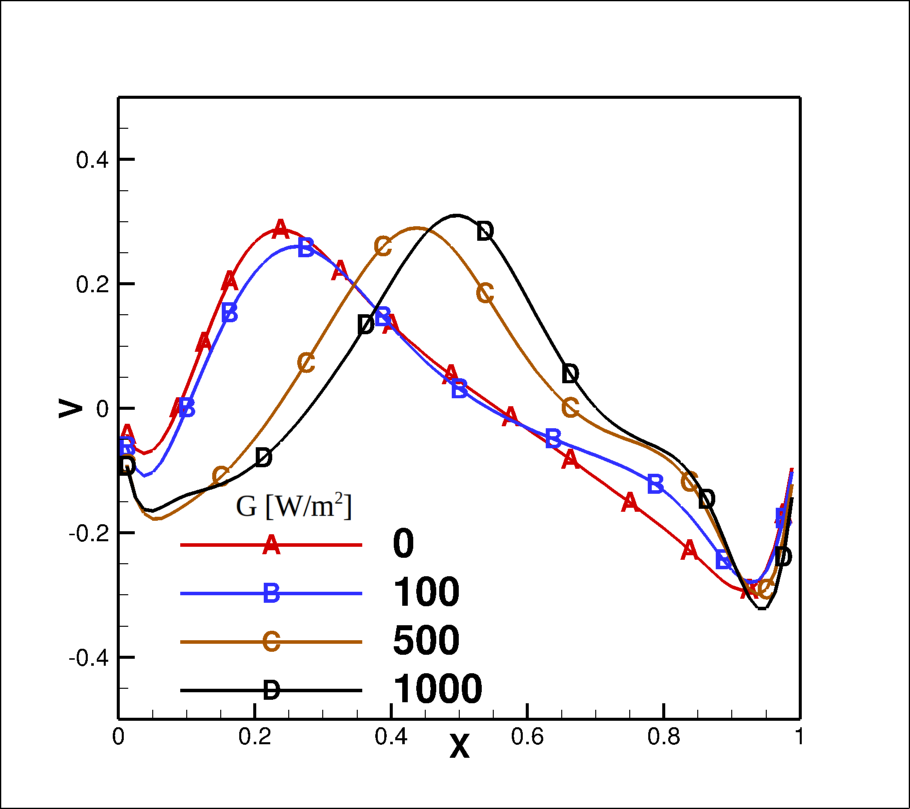}
    \caption{Variation of non-dimensional vertical velocity along the horizontal lines at the mid-height of the cavity for various values of irradiation on semitransparent window}
    \label{qcqr_vertV_mid}
 \end{figure}

The variation of non-dimensional vertical velocity in the horizontal direction at the mid height of the cavity is depicted in Fig. \ref{qcqr_vertV_mid}. The interesting fact to notice is that the maximum vertical velocity in downward direction increases for the left vortex upto irradiation values of 500 $W/m^2$ then slight decrements has been noticed for irradiation value 1000 $W/m^2$ while the maximum vertical velocity in downward direction remains almost constant for the right vortex for all values of irradiation. Furthermore, maximum vertical velocity in upward direction almost remains constant for all values of irradiation, while, its location keeps on shifting to right with increase in the value of irradiation. The value of this maximum non-dimensional vertical velocity in upward direction is 0.3 while maximum non-dimensional vertical velocity in downward direction is found right vortex and its values is 0.3.

\subsubsection{Heat transfer characteristics}
\begin{figure}[!t]
\begin{subfigure}{5cm}
    \centering\includegraphics[width=5cm]{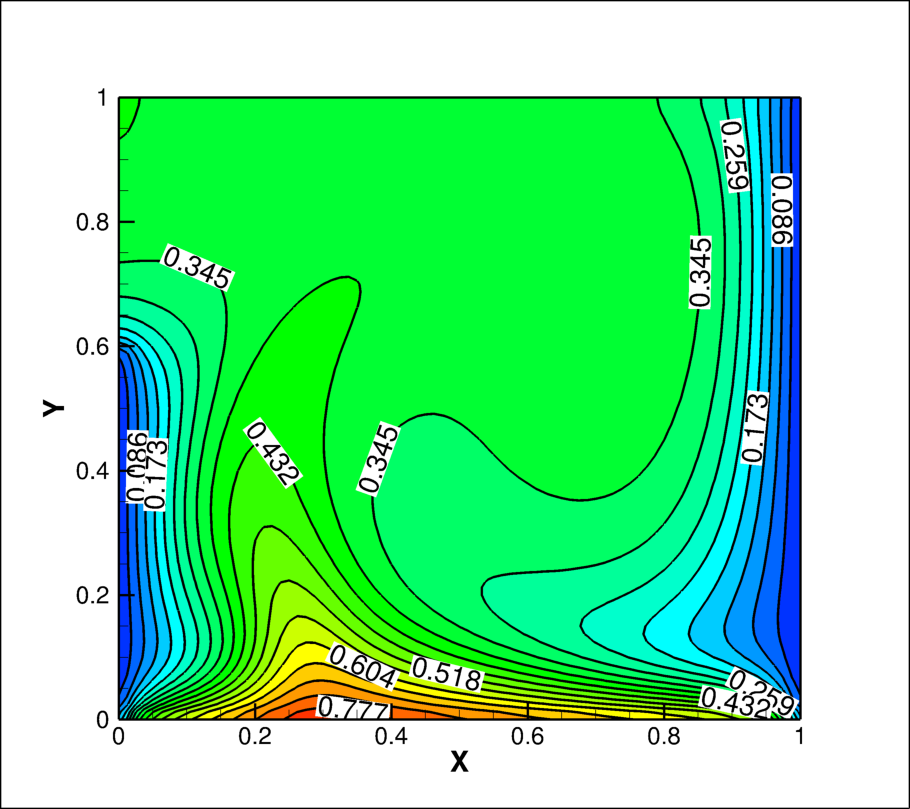}
    \caption{}
    \label{Temp_C1_N0}
  \end{subfigure}
   \begin{subfigure}{7cm}
    \centering\includegraphics[width=5cm]{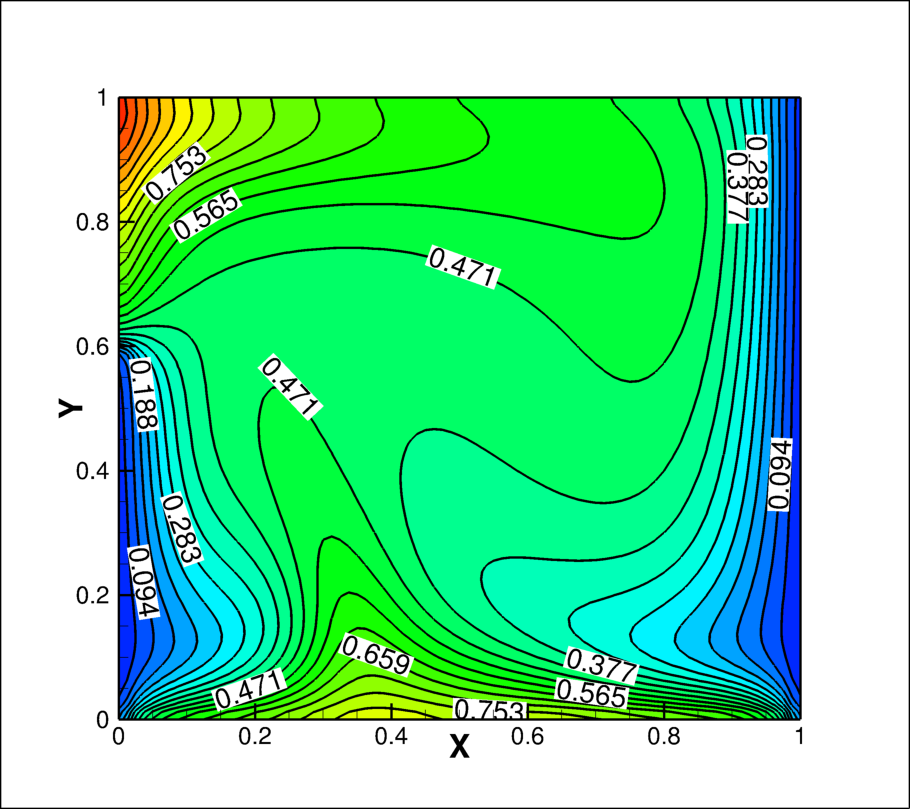}
    \caption{}
    \label{Temp_C1_N100}
  \end{subfigure}
  \begin{subfigure}{5cm}
    \centering\includegraphics[width=5cm]{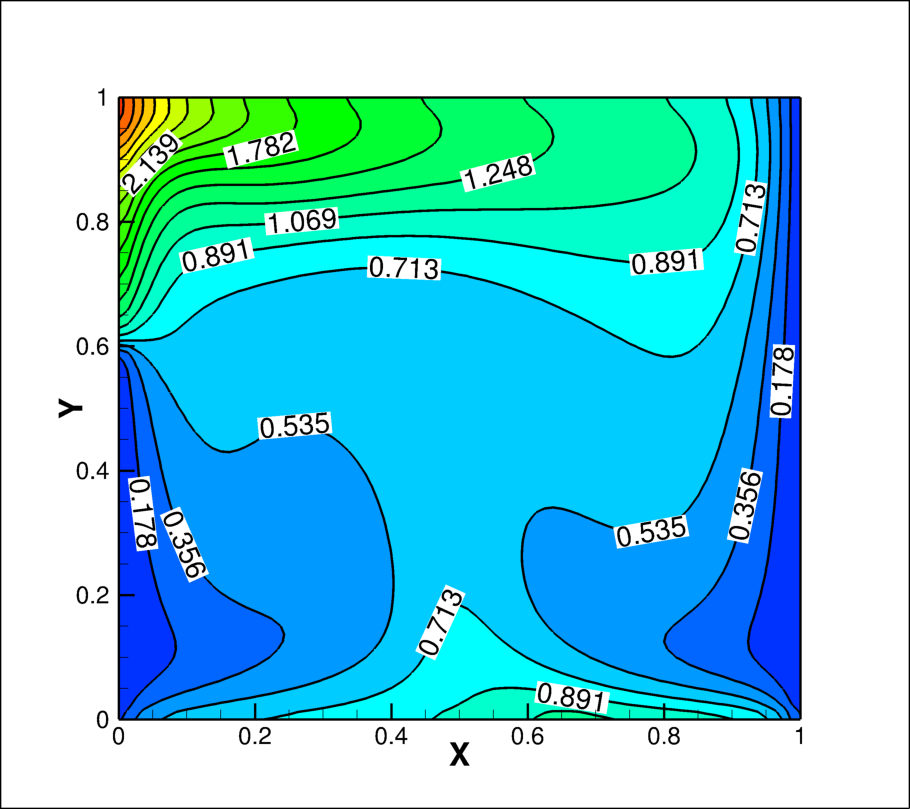}
    \caption{}
    \label{Temp_C1_N500}
  \end{subfigure}
   \begin{subfigure}{7cm}
    \centering\includegraphics[width=5cm]{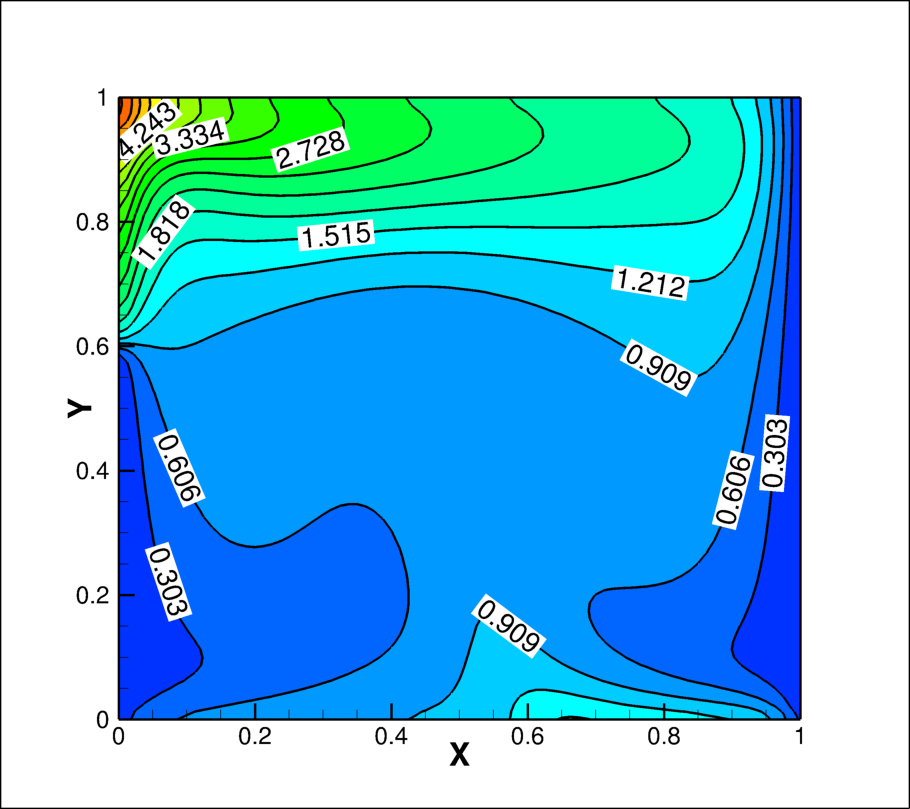}
    \caption{}
    \label{Temp_C1_N1000}
  \end{subfigure}
  \caption{Contours of non-dimensional isothermal for collimated irradiation of (a)G=0 (b)G=100 (c)G=500 and (d)G=1000 $W/m^2$ on semitransparent window}
\label{Temp_qcqr_collimated}
\end{figure}

The contours of non-dimensional temperature for various values of irradiation are shown in Fig. \ref{Temp_qcqr_collimated}. Although, the stream function contours show no difference between case A as B, (see Fig. \ref{SF_C_N0} and \ref{SF_C1_N0}) for irradiation value zero, the non-dimensional temperature contours for case B shows the difference at the upper part of the cavity near to semitransparent window, else it remain similar to case A. The clustering of isothermal lines are near to isothermal walls and dense clustered lines are visible near to bottom wall for irradiation value of 0 and 100 $W/m^2$. However, scenario changes drastically with the little increase in collimated irradiation. Now, the clustering of isothermal lines happens on semitransparent window and upper adiabatic wall, also, the clustering on isothermal and bottom walls have decreased. This phenomenon increases with increase of irradiation values. For high value of irradiation, the non-dimensional temperature is more uniform on lower part of the cavity than the upper part of cavity where high temperature exists inside the cavity unlike to case A. Table 5 shows the maximum non-dimensional temperature inside the cavity. The maximum non-dimensional temperature increases to 5.758 for irradiation value of 1000 $W/m^2$ which exist at the junction point of semitransparent and above adiabatic walls, unlike to case A which always occur on the bottom wall.

\begin{figure}[!b]
 \begin{subfigure}{5cm}
    \centering\includegraphics[width=5cm]{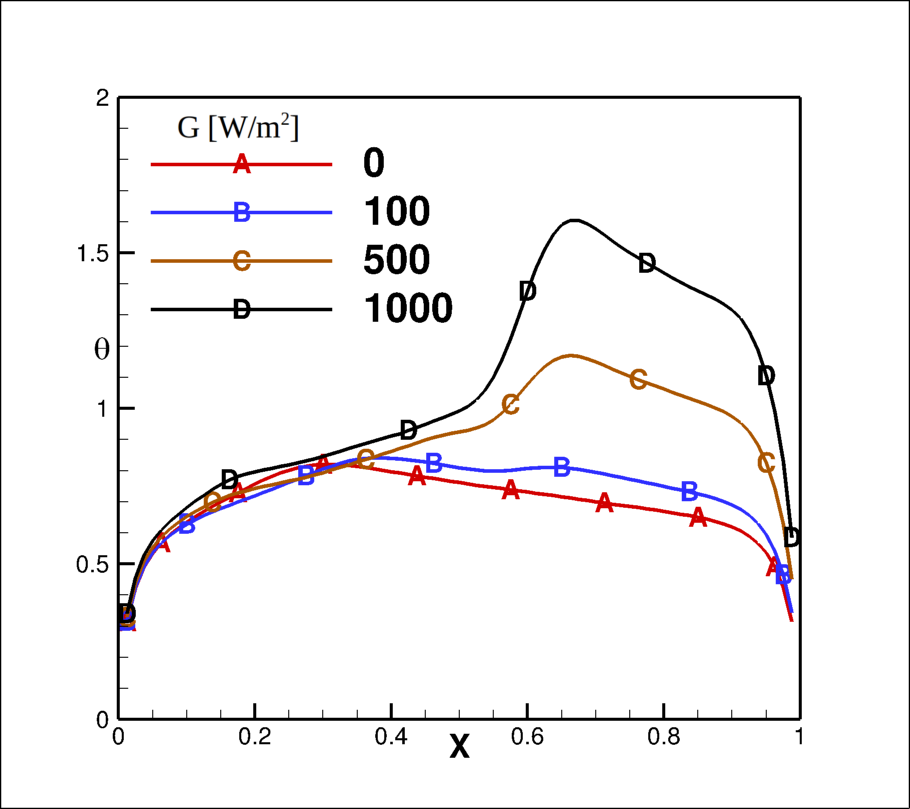}
    \caption{}
    \label{qcqr_bot_temp}
  \end{subfigure}
  \begin{subfigure}{7cm}
    \centering\includegraphics[width=5cm]{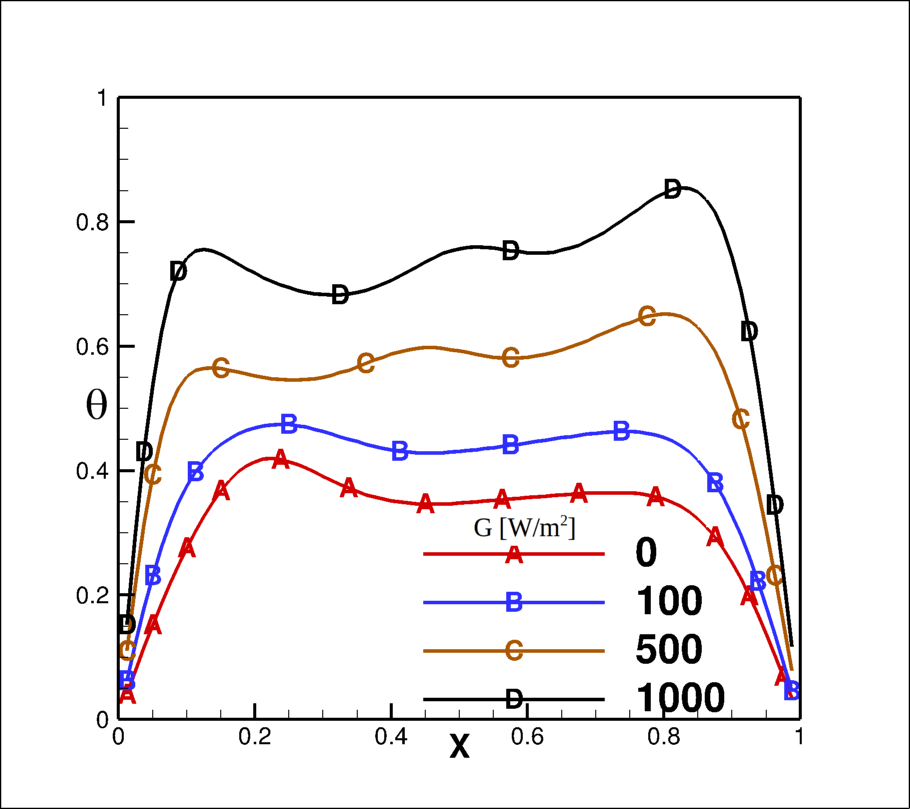}
    \caption{}
    \label{qcqr_mid_temp}
  \end{subfigure}
   \begin{subfigure}{5cm}
    \centering\includegraphics[width=5cm]{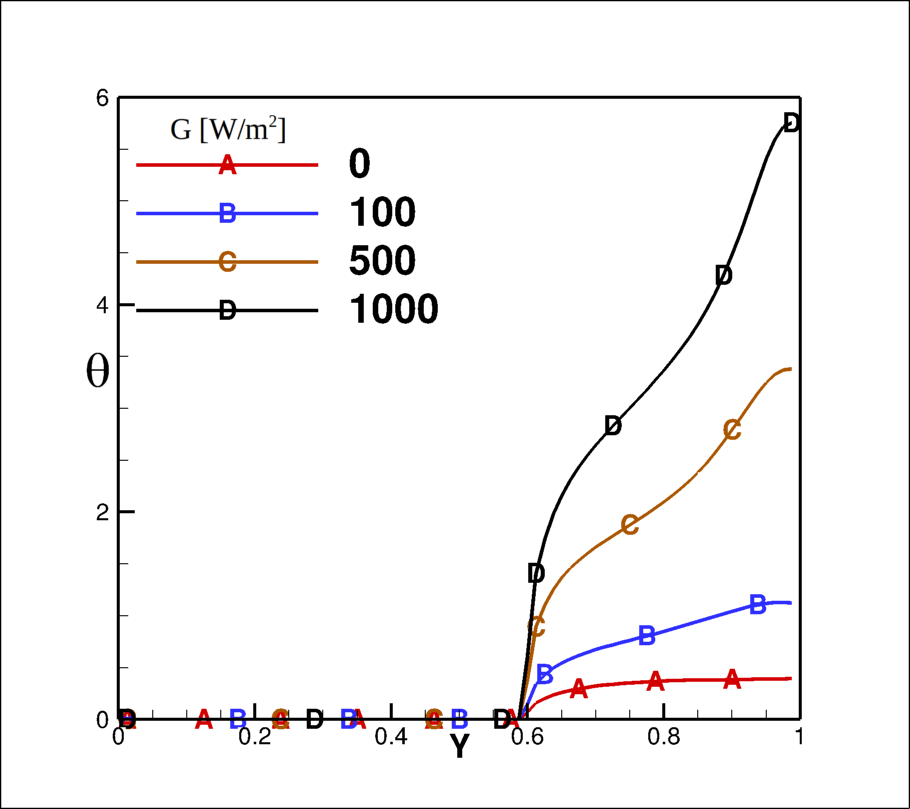}
    \caption{}
    \label{qcqr_left_temp}
  \end{subfigure}
  \begin{subfigure}{7cm}
    \centering\includegraphics[width=5cm]{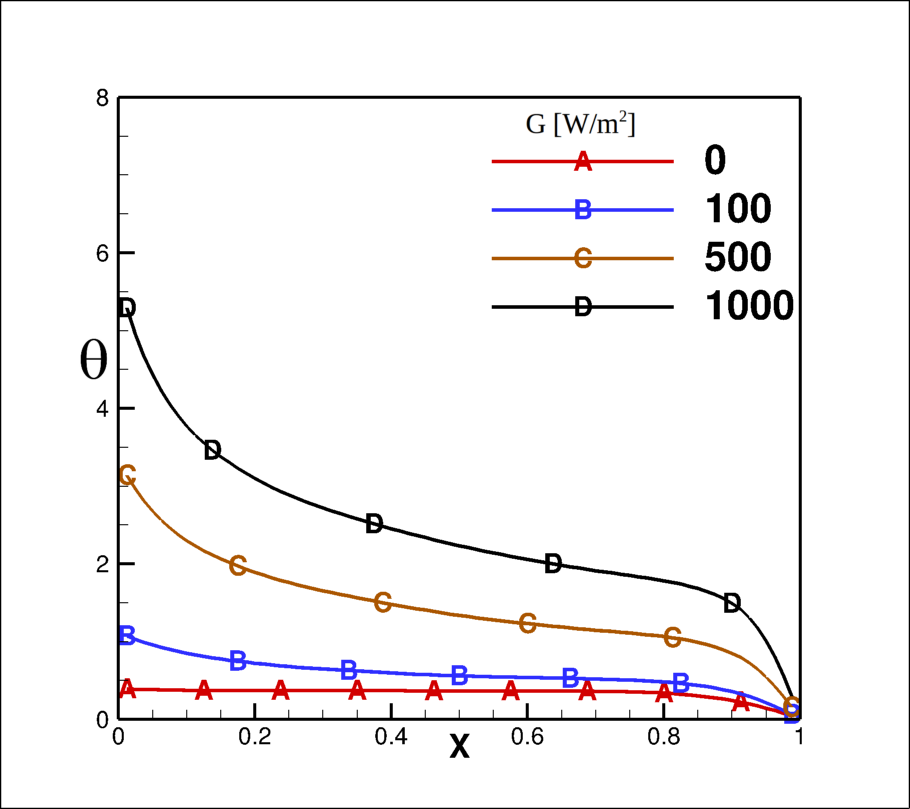}
    \caption{}
    \label{qcqr_top_temp}
  \end{subfigure}
  \caption{Variation of non-dimensional temperature at (a) bottom wall (b) horizontal line at mid-height of cavity (c) left wall and (d) top wall for various values of collimated irradiation on semitransparent window}
\label{qcqr_temp_walls}
\end{figure} 

The variation of non-dimensional temperature on the bottom wall, in horizontal direction at mid-height of the cavity, left and top walls are shown in Fig. \ref{qcqr_temp_walls} (a), (b), (c) and (d), respectively. The temperature increases slowly on the bottom wall from both the corners for lower values of irradiation. While spatial rate of increase of temperature is higher on the right corner of the wall for higher value of irradiation and sudden decrease in non-dimensional temperature is found just after the collimated beam strike zone. There is hardly any difference in temperature found near to left corner of the wall with irradiation values. Unlike to the temperature variation on the bottom wall, there is temperature difference in the horizontal direction for various values of irradiation at mid height of the cavity. The global maxima in the curve exists at the left for lower values of irradiation, whereas, there are three local maxima are found for higher values of irradiation and now, global maxima of temperature curve exists on the right. The non-dimensional temperature remains constant upto height of isothermal portion of the left wall, whereas, sudden increase in temperature has been noticed on the upper semitransparent window. The spatial rate of increase of temperature increases with increase of irradiation value. The continuity in the temperature curve on junction of left and top wall is also observed (Fig. \ref{qcqr_temp_walls} (c) and (d), afterwards temperature decreases along the horizontal direction on the top wall and matches temperature on right wall (isothermal wall). The spatial rate of decrease of temperature decreases with decrease in irradiation value.

\subsubsection{Nusselt number characteristics}

\begin{figure}[!t]
 \begin{subfigure}{5cm}
    \centering\includegraphics[width=5cm]{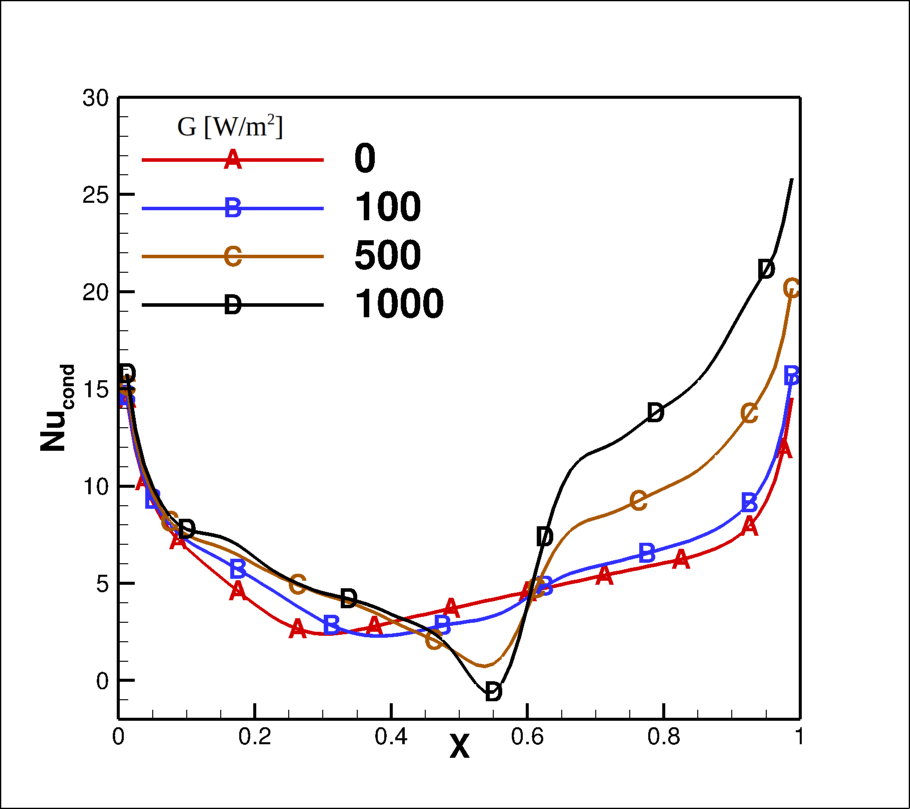}
    \caption{}
    \label{qcqr_bot_cond}
  \end{subfigure}
   \begin{subfigure}{5cm}
    \centering\includegraphics[width=5cm]{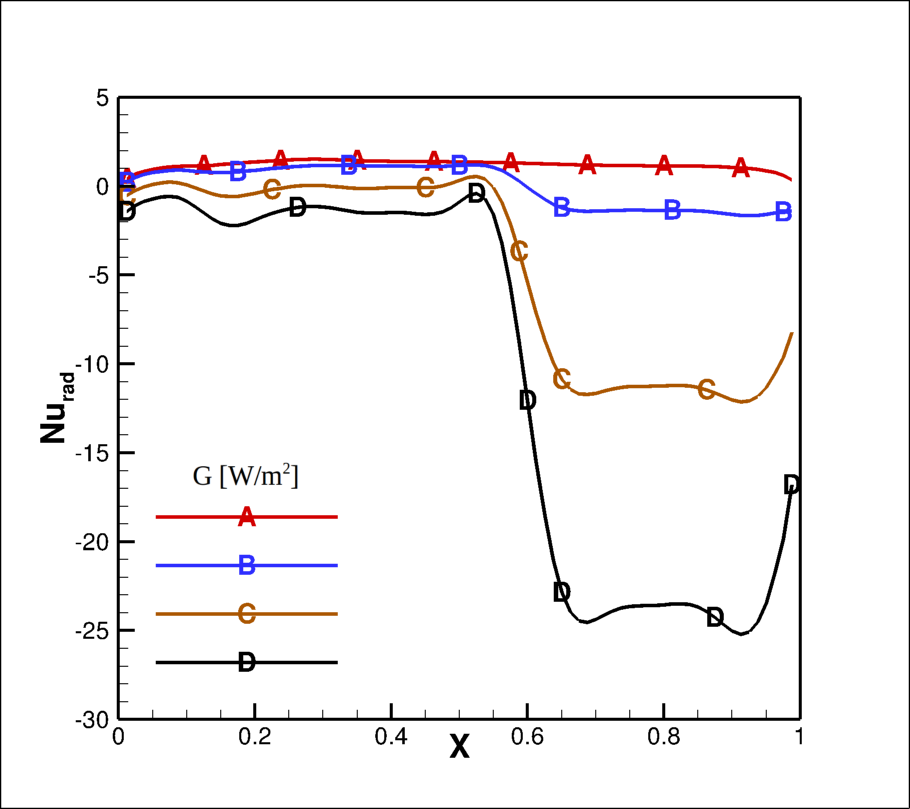}
    \caption{}
    \label{qcqr_bot_rad}
  \end{subfigure}
  \begin{subfigure}{12cm}
    \centering\includegraphics[width=5cm]{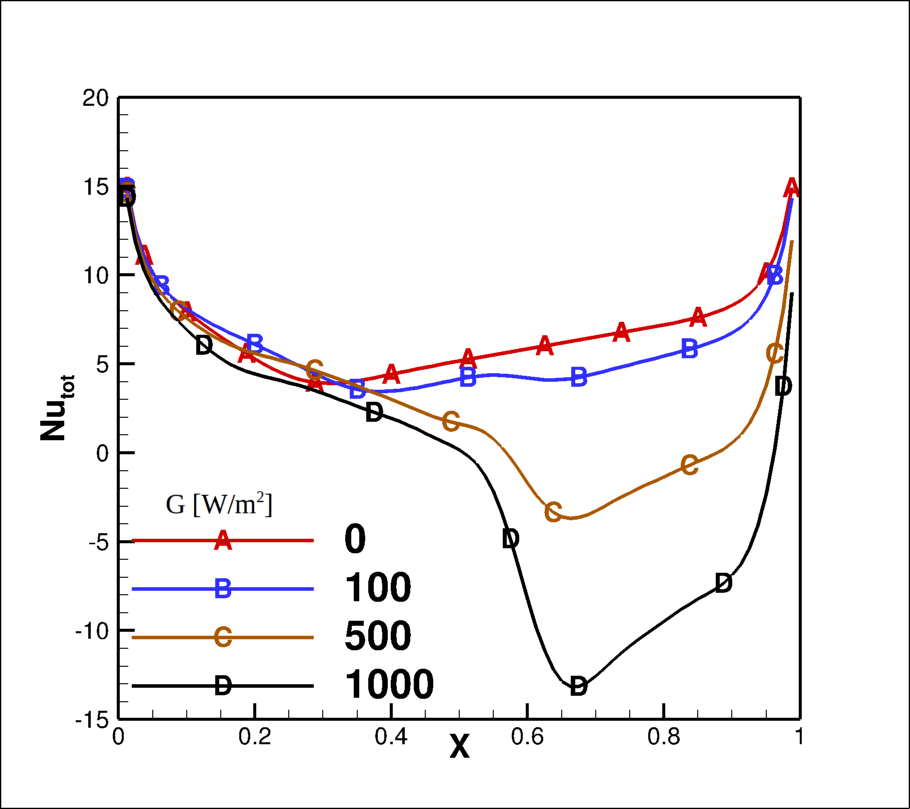}
    \caption{}
    \label{qcqr_bot_tot}
  \end{subfigure}
  \caption{Variation of (a) conduction (b) radiation and (c) total Nusselt number on the bottom wall for various values of collimated irradiation on semitransparent window}
\label{qcqr_bot_Nu}
\end{figure} 
 The conduction, radiation and total Nusselt number variation on the bottom wall are presented in the Fig. \ref{qcqr_bot_Nu} (a), (b) and (c), respectively for various values of irradiation. The conduction Nusselt number is almost same on both the ends of the bottom wall for the irradiation values of 0 and 100 $W/m^2$, whereas this trend changes with the increase in the irradiation value. The conduction Nusselt number remains almost same on the left corner of the bottom wall, and then slowly decreases to the minimum value at a non-dimensional distance of 0.3 from the left side wall for irradiation values 0 and 100 $W/m^2$, whereas this is minimum at a non-dimensional distance of 0.55 from the left corner for the irradiation values of 500 and 1000 $W/m^2$, and then there is sudden increase in the Nusselt number that reaches to the maximum value of 21 and 27 on right corner of the bottom wall for the irradiation values of 500 and 1000 $W/m^2$, respectively. The radiation Nusselt number is almost zero over the entire wall for the irradiation value 0 $W/m^2$, whereas, this trend is almost followed till the strike point, then there is increment
 observed in the radiation Nusselt number for any non-zero value of irradiation. This increment  is more pronounce for the irradiation values of 500 and 1000 $W/m^2$, where the most amount of energy leaves by radiation mode of heat transfer from the bottom wall. The total Nusselt number graphs follows the conduction Nusselt number graph till the strike point and then the radiation dominates over strike length of the beam.

\begin{figure}[!b]
 \begin{subfigure}{5cm}
    \centering\includegraphics[width=5cm]{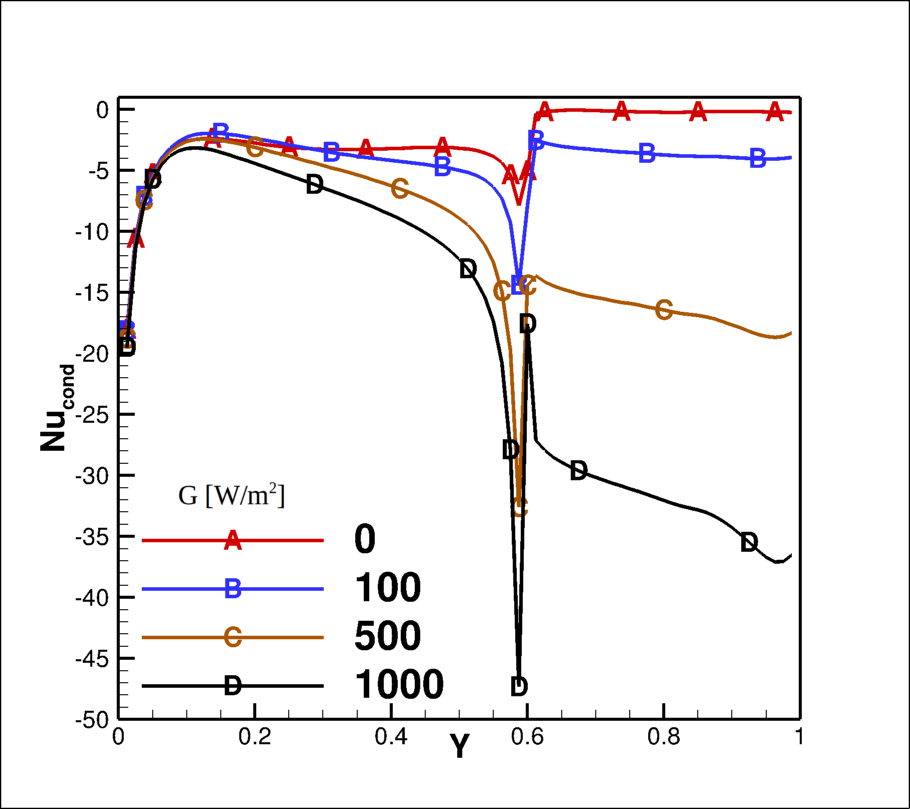}
    \caption{}
    \label{qcqr_left_cond}
  \end{subfigure}
   \begin{subfigure}{5cm}
    \centering\includegraphics[width=5cm]{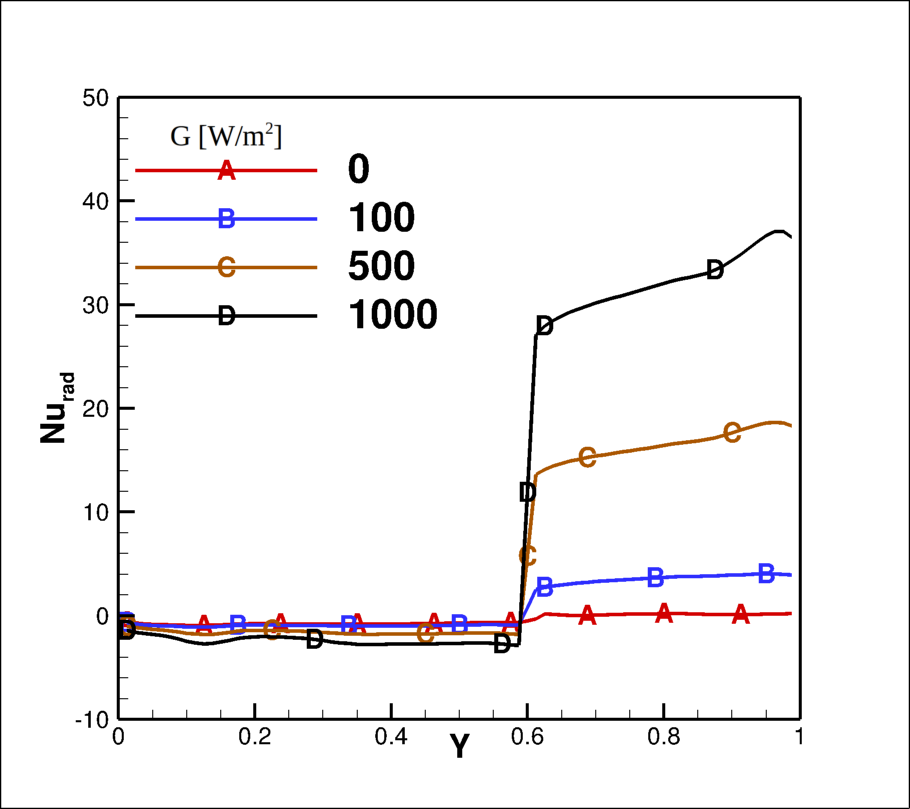}
    \caption{}
    \label{qcqr_left_rad}
  \end{subfigure}
  \begin{subfigure}{12cm}
    \centering\includegraphics[width=5cm]{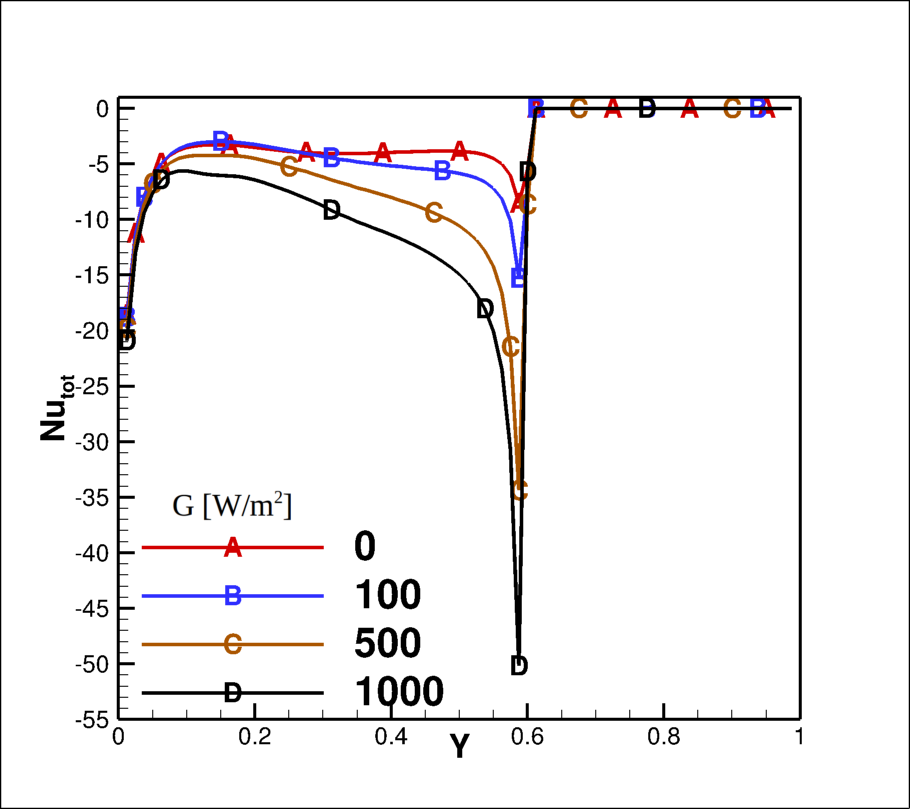}
    \caption{}
    \label{qcqr_left_tot}
  \end{subfigure}
  \caption{Variation of (a) conduction (b) radiation and (c) total Nusselt number on the left wall for various values of collimated irradiation on semitransparent window}
\label{qcqr_left_Nu}
\end{figure}

The variation of conduction, radiation and total Nusselt numbers on the vertical left wall which also contains the semitransparent wall are shown in Fig. \ref{qcqr_left_Nu}. The conduction Nusselt number decreases drastically with height and reaches to minimum  value 3 at non-dimensional height of 0.1; afterwards it remains almost constant on the isothermal left wall for lower values of irradiation, however, it starts increasing for higher value of irradiation.
There is sudden peak in the conduction Nusselt number near the junction point of isothermal wall and semitransparent wall where adiabatic condition is applied. Afterward, it decreases to further minimum value 0 and remain constant through out the height of semitransparent wall for irradiation value of zero and little increase in value for 100 $W/m^2$, whereas, similar peak with large value is observed in conduction Nusselt number near to junction point. After the peak the conduction Nusselt number, further increases slowly over the height of semitransparent wall and drastically increases for irradiation values of 500 and 100 $W/m^2$. The negative conductive Nusselt number indicates that the heat transferred from the system through conduction mode of heat transfer. The radiative Nusselt number (Fig. \ref{qcqr_left_rad}) is almost zero over the height of isothermal wall whereas it increases sudden over height of the semitransparent wall and increase with the irradiation value on semitransparent wall.

The total Nusselt number which is a linear sum of conduction and radiation Nusselt numbers is mostly governed by conduction Nusselt number on isothermal wall because of diffuse radiation. The total Nusselt number is zero on the semitransparent wall because of combinedly conductive and radiative adiabatic condition. The conduction Nusselt number is equal and opposite to radiative Nusselt number over the semitransparent wall. It can be verified by conduction and radiation Nusselt number (Fig. 19(a) and (b)) curves. Total Nusselt number also has a peak near the junction point of the isothermal and the semitransparent wall.

\begin{figure}[!t]
 \begin{subfigure}{5cm}
    \centering\includegraphics[width=5cm]{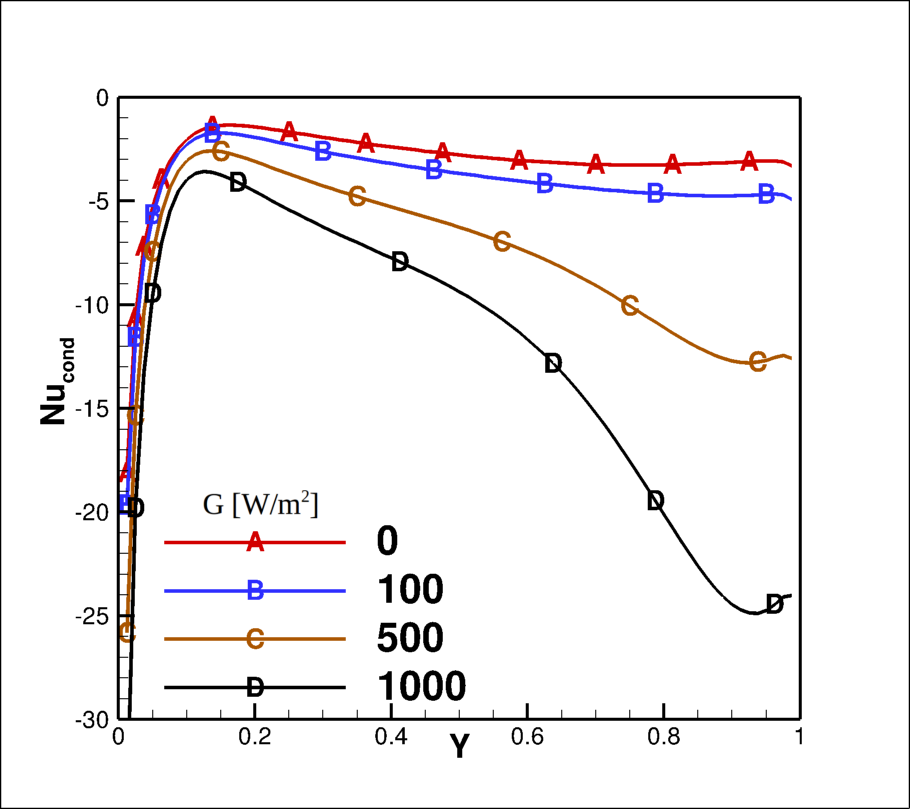}
    \caption{}
    \label{qcqr_right_cond}
  \end{subfigure}
   \begin{subfigure}{5cm}
    \centering\includegraphics[width=5cm]{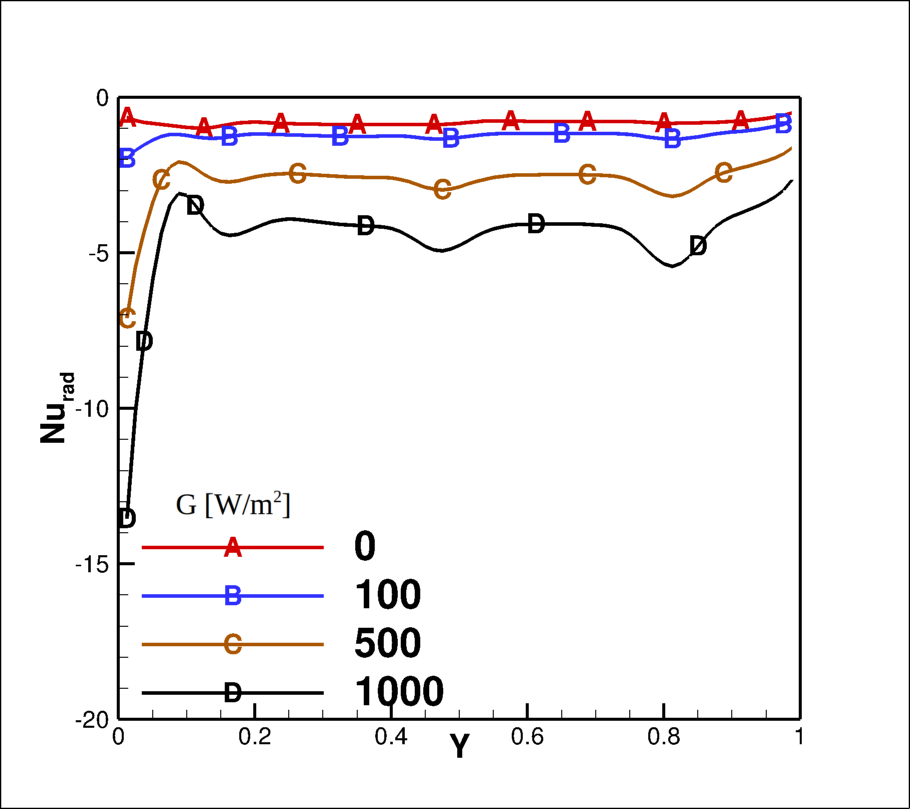}
    \caption{}
    \label{qcqr_right_rad}
  \end{subfigure}
  \begin{subfigure}{12cm}
    \centering\includegraphics[width=5cm]{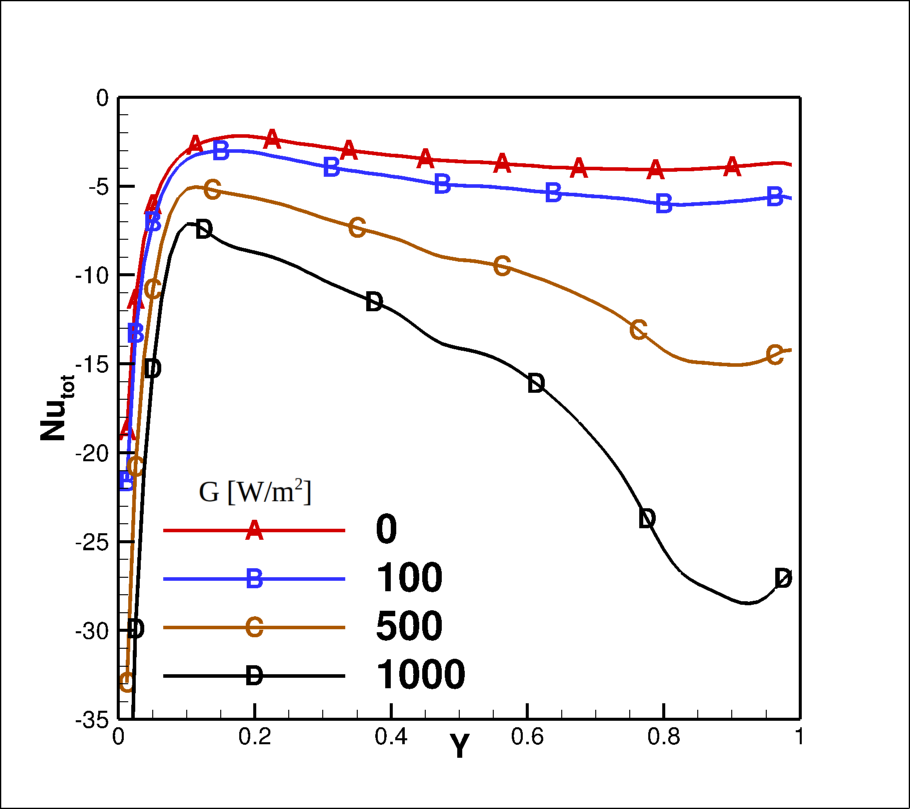}
    \caption{}
    \label{qcqr_right_tot}
  \end{subfigure}
  \caption{Variation of (a) conduction (b) radiation and (c) total Nusselt number on the right wall for various values of collimated irradiation on semitransparent window}
\label{qcqr_right_Nu}
\end{figure}

The conduction, radiation and total Nusselt number variations on the right wall are depicted in Fig. \ref{qcqr_right_Nu}(a), (b) and (c), respectively. The conduction Nusselt number decreases drastically upto non-dimensional height of 0.1 afterward it remain almost constant for irradiation value 0 $W/m^2$ while it increases slowly for lower value of irradiation and faster for higher value of irradiation over the rest height of the right wall. The radiative Nusselt number is very small for irradiation values of 0 and 100 $W/m^2$ while it has significant value for irradiation value of 500 and 1000 $W/m^2$. It shows that the heat transfer by diffuse radiation increases with the increase of collimated irradiation. Both conductive and radiative Nusselt numbers are negative which indicates that energy leaves the cavity by both the modes of heat transfer. The total Nusselt number is the linear sum of conduction and radiation Nusselt number, thus, conduction is being dominated mode of heat transfer in the present scenario, the total Nusselt number variation is mostly similar nature as of conduction Nusslet number only its values little is higher than the conduction Nusselt number due to additional of radiative Nusselt number.

\begin{landscape}
\begin{table}[!b]
\centering
\caption{Average Nusselt number on different walls for various values of irradiation for combinedly conductively and radiatively adiabatic condition on semitransparent wall}
\label{qcqr_avg_Nu_table}
\begin{tabular}{|c|c|c|c|c|c|c|c|c|c|c|c|}
\hline
\multirow{3}{*}{\begin{tabular}[c]{@{}c@{}}Irradiation\\ (G)\end{tabular}} & \multicolumn{3}{c|}{Bottom wall} & \multicolumn{5}{c|}{Left wall} & \multicolumn{3}{c|}{Right wall} \\ \cline{2-12} 
 & \multirow{2}{*}{Conduction} & \multirow{2}{*}{Radiation} & \multirow{2}{*}{Total} & \multicolumn{2}{c|}{Isothermal wall} & \multicolumn{2}{c|}{Semitransparent wall} &  & \multirow{2}{*}{Conduction} & \multirow{2}{*}{Radiation} & \multirow{2}{*}{Total} \\ \cline{5-9}
 &  &  &  & Conduction & Radiation & Conduction & Radiation & Total &  &  &  \\ \hline
0 & 5.393 & 1.202 & 6.595 & -2.328 & -0.454 & -0.06 & 0.06 & -2.782 & -2.982 & -0.802 & -3.784 \\ \hline
100 & 5.743 & 0.045 & 5.788 & -2.777 & -0.537 & -1.386 & 1.386 & -3.314 & -3.93 & -1.208 & -5.138 \\ \hline
500 & 7.047 & -4.451 & 2.596 & -4.226 & -0.864 & -6.378 & 6.378 & -5.09 & -7.523 & -2.656 & -10.173 \\ \hline
1000 & 8.734 & -10.013 & -1.279 & -5.641 & -1.267 & -12.495 & 12.495 & -6.908 & -12.351 & -4.384 & -16.735 \\ \hline
\end{tabular}
\end{table}
\end{landscape}

 Table 6 represents the average value of Nusselt number on different walls of cavity for various values of irradiation. The average conduction Nusselt number increases with the irradiation value while average radiation Nusselt number is positive for lower irradiation value then becomes negative making decreases in total Nusslet number with irradiation values on the bottom wall. At high irradiation value, the energy also starts leaving from the bottom wall. The left wall consists of two portions (1) lower isothermal wall (2) upper combinedly conductively and radiatively adiabatic wall. On lower isothermal wall both the conductive and radiative Nusselt numbers increases with the increase of irradiation value and similar on the upper semitransparent wall, but the conduction Nusselt number is same and opposite to radiative Nusselt number makes upper semitransparent wall combinedly conductively and radiatively adiabatic. The conduction, radiation and total Nusselt numbers behaviour on right side wall is similar to the lower part of the left side wall where the Nusselt numbers are negative and increases with increase of irradiation.

\section{Conclusions}
Two thermal adiabatic boundary conditions on semitransparent window have been investigated with diffuse/collimated irradiation on a natural convection in a cavity for various values of irradiation. These two thermal adiabatic boundary conditions arise based on the fact that the whether semitransparent window allows the energy leaves the cavity by radiation mode of heat transfer or not assuming being low thermal conductivity of semitransparent material, the energy does not leave by conduction mode of heat transfer. Thus, in this way, the semitransparent window may behave conductively adiabatic (case A) or combinedly conductively and radiatively adiabatic (case B) and the following conclusion are drawn
\begin{enumerate}
\item The dynamics of two vortices change with the change in irradiation values. There is increase in size and stream function values of the left vortex, while size of the right vortex decreases and stream function values remain constant with increase of irradiation value for case A, whereas planner flow exists at top of the cavity and left vortex remains in the lower part of the cavity for higher values of irradiation for case B.
\item Maximum downward and upward vertical velocities at mid height of the cavity is higher in case A than case B. 
\item The clustering of isothermal lines are near to the isothermal and bottom walls for all values of irradiation for case A while this clustering happen near to semitransparent window and upper adiabatic wall for higher values of irradiation for case B.
\item The non-dimensional temperature rises multi-fold inside the cavity for case B than to case A and this high rise in temperature exist near to junction point of semitransparent window and upper adiabatic wall for case B while at strike length of collimated beam on bottom wall for case A.
\item The diffuse radiation has less influence in the present problem. The Nusselt number is majorly dominated by conduction Nusselt number upto the non-strike length of collimated irradiation, afterward it is dominated by collimated beam radiation..
\item The total Nusselt number decreases with increase of irradiation on the bottom wall while it increases on both the vertical walls for both the cases. The Nusselt number becomes negative on bottom wall for higher values of irradiation for case B while it remains positive for case A.

\end{enumerate}

\section*{Acknowledgements}
The authors greatly acknowledge the financial support provided by Science and Engineering Research Board (SERB) (Statutory Body of the Government of India) via Grant. No: ECR/2015/000327 to carry out the present work.

\section*{Declaration of interest}
The authors declare that they have no known financial interests or personal relationships that could have appeared to influence the work reported in this paper.



\bibliography{mybibfile}

\begin{thebibliography}{10}
\expandafter\ifx\csname url\endcsname\relax
  \def\url#1{\texttt{#1}}\fi
\expandafter\ifx\csname urlprefix\endcsname\relax\def\urlprefix{URL }\fi
\expandafter\ifx\csname href\endcsname\relax
  \def\href#1#2{#2} \def\path#1{#1}\fi

\bibitem{Torrance}
{K.E. Torrance, J.A. Rockett}, Numerical study of natural convection in an
  enclosure with localized heating from below-creeping flow to the onset of
  laminar instability, J.Fluid Mech, part 1 36 (1969) 33--54.
\newblock \href {http://dx.doi.org/10.1017/S0022112069001492}
  {\path{doi:10.1017/S0022112069001492}}.

\bibitem{Calcagani}
{B. Calcagani, F. Marsili, M. Paroncini}, Natural convective heat transfer in
  square enclosures heated from below, App. Thermal engineering 25 (2005)
  2522--2531.
\newblock \href {http://dx.doi.org/10.1016/j.applthermaleng.2004.11.032}
  {\path{doi:10.1016/j.applthermaleng.2004.11.032}}.

\bibitem{Ganzorolli}
{M.M. Ganzorolli, L.F. Milanez}, Natural convection in rectangular enclosures
  heated from below and symmetrically cooled from the sides, Int.J. Heat mass
  Transfer, (6) 36 (1995) 1063--1073.
\newblock \href {http://dx.doi.org/10.1016/0017-9310(94)00217-J}
  {\path{doi:10.1016/0017-9310(94)00217-J}}.

\bibitem{Aydin}
{Orhan Aydin, Wen-Hei Yang}, Natural convection in enclosures with localized
  heating from below and symmetrical cooling from sides, Int. J of Num. Methods
  for Heat \& Fluid flow, No.5 10 (2000) 518--529.
\newblock \href {http://dx.doi.org/10.1108/09615530010338196}
  {\path{doi:10.1108/09615530010338196}}.

\bibitem{Sathiyamoorthy}
{M. Sathiyamoorthy, Tanmay Basak, S. Roy, I. Pop}, Steady natural convection
  flows in a square cavity with linearly heated side wall(s), Int. J. of Heat
  and Mass Transfer 50 (2007) 766--775.
\newblock \href {http://dx.doi.org/10.1016/j.ijheatmasstransfer.2006.06.019}
  {\path{doi:10.1016/j.ijheatmasstransfer.2006.06.019}}.

\bibitem{Acharya}
{S. Acharya, R.J. Goldstein}, Natural convection in an externally heated
  vertical or inclined square box containing internal energy sources, J. Of
  Heat Transfer 107 (1985) 855--866.
\newblock \href {http://dx.doi.org/10.1115/1.3247514}
  {\path{doi:10.1115/1.3247514}}.

\bibitem{Osman}
{Osman Turan, Robert J. Poole, Nilanjan Chakraborty}, Boundary condition
  effects on natural convection of bingham fluids in a square enclosure with
  differentially heated, Computational Thermal Sciences 4(1) (2012) 77--97.
\newblock \href {http://dx.doi.org/10.1615/ComputThermalScien.2012004759}
  {\path{doi:10.1615/ComputThermalScien.2012004759}}.

\bibitem{Rahimi}
{Alireza Rahimi, Ali Dehghan Saee, Abbas Kasaeipoor, DEmad, Hasani Malekshah},
  A comprehensive review on natural convection flow and heat transfer the most
  practical geometries for engineering applications, Int. J. of Numerical
  Methods for Heat and Fluid Flow 29 Issue: 3 (2019) 834--877.
\newblock \href {http://dx.doi.org/10.1108/HFF-06-2018-0272}
  {\path{doi:10.1108/HFF-06-2018-0272}}.

\bibitem{Das}
{Debayan Das, Monisha Roy, Tanmay Basak}, Studies on natural convection within
  enclosures of various (non- square) shapes-a review, Int. J. of Heat and Mass
  Transfer 106 (2017) 356--406.
\newblock \href {http://dx.doi.org/10.1016/j.ijheatmasstransfer.2016.08.034}
  {\path{doi:10.1016/j.ijheatmasstransfer.2016.08.034}}.

\bibitem{Bittagopal}
{B. Mondal, S. C. Mishra}, Simulation of natural convection in the presence of
  volumetric radiation using the lattice boltzmann method, Num.Heat
  Transfer,part-A 55 (2009) 18--41.
\newblock \href {http://dx.doi.org/10.1080/10407780802603121}
  {\path{doi:10.1080/10407780802603121}}.

\bibitem{Liu}
{L. H. Liu, H. C. Zhang, H. P. Tan}, Monte carlo discrete curved ray-tracing
  method for radiative transfer in an absorbing-emitting semi-transparent slab
  with variable spatial refractive index, J.of Quantative Spectroscopy and
  Radiative Transfer 84 (2004) 357--362.
\newblock \href {http://dx.doi.org/10.1016/S0022-4073(03)00186-9}
  {\path{doi:10.1016/S0022-4073(03)00186-9}}.

\bibitem{Kumar}
{P. Kumar, V. Eswaran}, The effect of radiation on natural convection in
  slanted cavities of angle $45^0$ and $60^0$, Int.J.Thermal Science 67 (2013)
  96--106.
\newblock \href {http://dx.doi.org/10.1016/j.ijthermalsci.2012.12.009}
  {\path{doi:10.1016/j.ijthermalsci.2012.12.009}}.

\bibitem{Lari}
{K. Lari, M. Baneshi, S. G. Nassab, A. Komiya, S. Maruyama}, Combined heat
  transfer of radiation and natural convection in a square cavity containing
  participating gases, Int.J.Heat Mass Transfer 54 (2011) 5087--5099.
\newblock \href {http://dx.doi.org/10.1016/j.ijheatmasstransfer.2011.07.026}
  {\path{doi:10.1016/j.ijheatmasstransfer.2011.07.026}}.

\bibitem{Mezrhab}
{A. Mezrhab, H.Bouali, H. Amaoui, M. Bouzidi}, Compuations of combined natural
  convection and radiation heat transfer in a cavity having a square body at
  its centre, App.Energy 83 (2006) 1004--1023.
\newblock \href {http://dx.doi.org/10.1016/j.apenergy.2005.09.006}
  {\path{doi:10.1016/j.apenergy.2005.09.006}}.

\bibitem{Hua}
{Hua Sun, Eric Chenier, Guy Lauriat}, Effect of surface radiation on the
  breakdown of steady natural convection flows in a square, air-filled cavity
  containing a centred inner body, App. Thermal Engineering 31 (2011)
  252--1262.
\newblock \href {http://dx.doi.org/10.1016/j.applthermaleng.2010.12.028}
  {\path{doi:10.1016/j.applthermaleng.2010.12.028}}.

\bibitem{Mukul}
{Mukul Paramanda, Salman Khan, Amaresh Dalal, Ganesh Natarajan}, Critical
  assessment of numerical alogorithms for convective-radiative heat transfer in
  enclosures with different geometries, Int.J. of Heat and Mass Transfer 108
  (2017) 627--644.
\newblock \href {http://dx.doi.org/10.1016/j.ijheatmasstransfer.2016.12.033}
  {\path{doi:10.1016/j.ijheatmasstransfer.2016.12.033}}.

\bibitem{Xu}
{Xu Xu, Gonggang Sun, Zitao Yu, Yacai Hu, Liwu Fan, Kefa Cen}, Numerical
  investigation of laminar natural convective heat transfer from a horizontal
  triangular cylinder to its concentric cylindrical enclosure, Int. J. Heat and
  Mass Transfer 52 (2009) 3176–3186.
\newblock \href {http://dx.doi.org/10.1016/j.ijheatmasstransfer.2009.01.026}
  {\path{doi:10.1016/j.ijheatmasstransfer.2009.01.026}}.

\bibitem{Chai}
{John C. Chai, HaeOk S. Lee, Suhas V. Patankar}, Ray effect and false
  scattering in the discrete ordinates method, Numerical Heat Transfer, Part B
  24 (1993) 373--389.
\newblock \href {http://dx.doi.org/10.1080/10407799308955899}
  {\path{doi:10.1080/10407799308955899}}.

\bibitem{Raithby}
{G. D. Raithby, E. H. Chui}, A finite-volume method for predicting a radiant
  heat transfer in enclosures with participating media, Trans. of the ASME,
  Journal of Heat Transfer 112 (1990) 415–423.
\newblock \href {http://dx.doi.org/10.1115/1.2910394}
  {\path{doi:10.1115/1.2910394}}.

\bibitem{Chui}
{E. H. Chui, G. D. Raithby}, Computation of radiant heat transfer on a
  nonorthogonal mesh using the finite-volume method, Numerical Heat Transfer,
  PartB: Fundamental 23 (1993) 269–288.
\newblock \href {http://dx.doi.org/10.1080/10407799308914901}
  {\path{doi:10.1080/10407799308914901}}.

\bibitem{Yujia}
{Yujia Sun, Xiaobing Zhang, J. R. Howell}, Assessment of different radiative
  transfer equation solvers for combined natural convection and radiation heat
  transfer problems, J. of Quantitative Spectroscopy and Radiative Transfer 194
  (2017) 31--46.
\newblock \href {http://dx.doi.org/10.1016/j.jqsrt.2017.03.022}
  {\path{doi:10.1016/j.jqsrt.2017.03.022}}.

\bibitem{Hakan}
{Hakan Karatas, Taner Derbentli}, Natural convection and radiation in
  rectangular cavities with one active vertical wall, Int. J of Thermal
  sciences 123 (2018) 129 -- 139.
\newblock \href {http://dx.doi.org/10.1016/j.ijthermalsci.2017.09.006}
  {\path{doi:10.1016/j.ijthermalsci.2017.09.006}}.

\bibitem{Webb}
{B. W. Webb, R. Viskanta}, Radiation-induced buoyancy driven flow in
  rectangular enclosures: Experiment and analysis, J.of Heat Transfer 109
  (1987) 427--433.
\newblock \href {http://dx.doi.org/10.1115/1.3248099}
  {\path{doi:10.1115/1.3248099}}.

\bibitem{Anand}
{N. Anand Krishna, S. C. Mishra}, Discrete transfer method applied to radiative
  transfer in variable refractive index, J.of Quantative Spectroscopy and
  Radiative Transfer 102 (2006) 432--440.
\newblock \href {http://dx.doi.org/10.1016/j.jqsrt.2006.02.024}
  {\path{doi:10.1016/j.jqsrt.2006.02.024}}.

\bibitem{Ben}
{P. Ben Abdallah, V. Le Dez}, Radiative flux field inside an absorbing-emitting
  semi transparent slab with a variable refractive index at radiative
  conductive coupling, J.of Quantative Spectroscopy and Radiative Transfer
  {67(2)} (2000) 125--137.
\newblock \href {http://dx.doi.org/10.1016/S0022-4073(99)00200-9}
  {\path{doi:10.1016/S0022-4073(99)00200-9}}.

\bibitem{patankar}
{S. V. Patankar}, Numerical heat transfer and fluid flow, Hemisphere Publishing
  Corporation, 1980.

\bibitem{Moukalled}
{F. Moukalled, L. Mangani, M. Darwish}, The Finite Volume Method in
  Computational Fluid Dynamics: An Advanced Introduction with OpenFOAM and
  Matlab, Springer International Publishing, 2016.

\bibitem{RAD19}
{Ankur Garg, G Chanakya, Pradeep Kumar}, Numerical error estimation in finite
  volume method for radiative transfer equation for collimated irradiation, in:
  Proceedings of the 9th International Symposium on Radiative Transfer, RAD-19,
  June 3-7, 2019, Athens, Greece, 2019.

\bibitem{Aswatha}
{Aswatha, C. J. Gangadhara Gowda, S. N. Sridhara, K. N. Seetharamu}, Effect of
  different thermal boundary conditions at bottom wall on natural convection in
  cavities, J. of Engineering Science and Technology 6 (2011) 109 -- 130.

\bibitem{openfoam2017open}
OpenFOAM, The open source cfd toolbox: User guide, openfoam v1706 (2017).

\bibitem{Basak}
{Tanmay Basak, S. Roy, A.R. Balakrishnan}, Effects of thermal boundary
  conditions on natural convection flows within a square cavity, Int. J of Heat
  and Mass Transfer 4525–4535 49 (2006) 4525--4535.
\newblock \href {http://dx.doi.org/10.1016/j.ijheatmasstransfer.2006.05.015}
  {\path{doi:10.1016/j.ijheatmasstransfer.2006.05.015}}.

\end{thebibliography}

\end{document}